\documentclass[10pt,conference]{IEEEtran}
\IEEEoverridecommandlockouts
\usepackage{tikz}
\usepackage{pgfplots}
\usepackage{pgfplotstable}
\pgfplotsset{compat=1.18}
\usetikzlibrary{patterns,arrows.meta,positioning,fit,backgrounds,calc,decorations.pathreplacing}
\providecommand{\cmark}{\textcolor{green!55!black}{\ding{51}}}

\colorlet{cAgg}{blue!70!black}
\colorlet{cRec}{orange!80!black}
\colorlet{cRel}{green!55!black}
\newcommand{\dfg}{\tikz[baseline=-0.5ex]{
  \node[draw=cAgg,fill=cAgg!12,rounded corners=0.3pt,minimum width=2.7mm,
        minimum height=2.5mm,inner sep=0](b){};
  \fill[cAgg!75] (b.north west) rectangle ([yshift=-0.6mm]b.north east);}}
\newcommand{\dff}{\dfg\kern0.4pt\dfg\kern0.4pt\dfg}
\newcommand{\aggfn}[1]{\textcolor{cAgg}{\textbf{\textsf{#1}}}}
\newcommand{\recfn}[1]{\textcolor{cRec}{\textbf{\textsf{#1}}}}
\newcommand{\relsym}[1]{\textcolor{cRel}{$\mathbf{#1}$}}

\usepackage{siunitx}
\usepackage{multirow, listings}
\usepackage{pifont}
\usepackage{colortbl}
\usepackage{subcaption}
\usepackage{amsmath}
\usepackage{amssymb}
\usepackage{graphicx, subcaption}
\usepackage{soul}
\usepackage{balance}
\usepackage{makecell}
\usepackage{algorithm}
\usepackage{microtype}
\usepackage{xcolor}
\usepackage{caption}
\usepackage{subcaption}
\usepackage{hyperref}
\usepackage{array}
\usepackage{booktabs}
\usepackage{tabularx}
\usepackage{makecell,array}
\usepackage{hhline}
\usepackage{longtable}
\usepackage{arydshln}
\let\origtexttt\texttt
\renewcommand{\texttt}[1]{\origtexttt{\fontsize{0.9em}{0.95em}\selectfont #1}}

\usepackage{xspace}
\usepackage{comment}

\pgfplotsset{
  cov style/.style={
    width=\linewidth, height=4.6cm,
    xlabel={\footnotesize Test invocations},
    legend style={font=\scriptsize, at={(0.98,0.08)}, anchor=south east},
    grid=major, grid style={dashed, gray!20},
    xmin=0, xmax=510,
    xtick={0,100,200,300,400,500},
    xticklabel style={font=\scriptsize},
    yticklabel style={font=\scriptsize},
    ylabel style={font=\footnotesize},
  },
  risk style/.style={
    width=\linewidth, height=4.0cm,
    xlabel={\footnotesize Test invocations},
    legend style={font=\scriptsize, at={(0.98,0.98)}, anchor=north east},
    grid=major, grid style={dashed, gray!20},
    xmin=21, xmax=510,
    xtick={50,150,250,350,450},
    xticklabel style={font=\scriptsize},
    yticklabel style={font=\scriptsize},
    ylabel style={font=\footnotesize},
  },
}
\usepackage{listings}
\usepackage{xcolor}
\usepackage{kotex}
\usepackage{enumitem}

\definecolor{claimcolor}{rgb}{0.0, 0.4, 0.0}
\definecolor{riskcolor}{rgb}{0.7, 0.1, 0.1}

\newcommand{\tool}{\textsc{DualVeri}\xspace}

\usepackage[normalem]{ulem}

\newcommand{\simplebox}[2]{%
  \par\noindent
  \setlength{\fboxsep}{4pt}%
  \setlength{\fboxrule}{1.4pt}%
  \fcolorbox{black}{#1}{%
    \parbox{\dimexpr\linewidth-2\fboxsep-2\fboxrule\relax}{#2}%
  }%
  \par\vspace{4pt}%
}

\lstdefinelanguage{lean4}{
  keywords={theorem,def,lemma,by,have,let,show,exact,apply,intro,induction,cases,simp,rfl,sorry,where,structure,class,instance,fun,match,if,then,else,return,do,inductive,forall,exists},
  keywordstyle=\color{blue}\bfseries,
  comment=[l]{--},
  commentstyle=\color{gray}\itshape,
  string=[b]",
  stringstyle=\color{teal},
  basicstyle=\ttfamily\footnotesize,
  columns=fullflexible,
  keepspaces=true,
  showstringspaces=false,
  breaklines=true,
  frame=single,
  numbers=left,
  numberstyle=\ttfamily\tiny,
  xleftmargin=1.2em,
  literate=%
    {→}{{->}}2 {←}{{<-}}2 {↦}{{$\mapsto$}}1 {·}{{$\cdot$}}1 {…}{{\ldots}}1
    {⟹}{{$\Longrightarrow$}}1 {⟺}{{$\Longleftrightarrow$}}1
    {≃}{{$\simeq$}}1 {≈}{{$\approx$}}1 {≋}{{$\approxeq$}}1
    {×}{{$\times$}}1 {∘}{{$\circ$}}1
    {α}{{$\alpha$}}1 {β}{{$\beta$}}1 {γ}{{$\gamma$}}1 {δ}{{$\delta$}}1
    {ε}{{$\varepsilon$}}1 {ρ}{{$\rho$}}1 {σ}{{$\sigma$}}1 {τ}{{$\tau$}}1
    {φ}{{$\varphi$}}1 {ψ}{{$\psi$}}1 {ω}{{$\omega$}}1 {λ}{{$\lambda$}}1
    {Σ}{{$\Sigma$}}1 {Π}{{$\Pi$}}1
    {∀}{{$\forall$}}1 {∃}{{$\exists$}}1
    {≤}{{$\leq$}}1 {≥}{{$\geq$}}1 {≠}{{$\neq$}}1 {≡}{{$\equiv$}}1
    {∈}{{$\in$}}1 {∉}{{$\notin$}}1 {⊆}{{$\subseteq$}}1 {⊂}{{$\subset$}}1
    {∧}{{$\wedge$}}1 {∨}{{$\vee$}}1 {¬}{{$\neg$}}1
    {ℕ}{{$\mathbb{N}$}}1 {ℤ}{{$\mathbb{Z}$}}1 {ℝ}{{$\mathbb{R}$}}1
    {⊢}{{$\vdash$}}1 {⊤}{{$\top$}}1 {⊥}{{$\bot$}}1
    {⊕}{{$\oplus$}}1 {⊗}{{$\otimes$}}1 {⊖}{{$\ominus$}}1 {⊙}{{$\odot$}}1
    {₀}{{$_0$}}1 {₁}{{$_1$}}1 {₂}{{$_2$}}1 {₃}{{$_3$}}1 {₄}{{$_4$}}1
    {₅}{{$_5$}}1 {₆}{{$_6$}}1 {₇}{{$_7$}}1 {₈}{{$_8$}}1 {₉}{{$_9$}}1
    {⁰}{{$^0$}}1 {¹}{{$^1$}}1 {²}{{$^2$}}1 {³}{{$^3$}}1 {⁴}{{$^4$}}1
    {⁵}{{$^5$}}1 {⁶}{{$^6$}}1 {⁷}{{$^7$}}1 {⁸}{{$^8$}}1 {⁹}{{$^9$}}1,
}

\lstdefinestyle{pyspark}{
  language=Python,
  basicstyle=\ttfamily\footnotesize,
  columns=fullflexible,
  keepspaces=true,
  showstringspaces=false,
  breaklines=true,
  frame=single,
  numbers=left,
  numberstyle=\ttfamily\tiny,
  xleftmargin=1.2em,
  keywordstyle=\color{blue}\bfseries,
  commentstyle=\color{gray}\itshape,
  stringstyle=\color{teal},
  identifierstyle=\color{black},
  emph={spark,createDataFrame,groupBy,agg,alias,first,filter,limit,count,sum},
  emphstyle=\color{purple}
}

\usepackage{listings}

\lstdefinestyle{prompt}{
  basicstyle=\ttfamily\footnotesize,
  breaklines=true,
  breakatwhitespace=false,
  columns=fullflexible,
  keepspaces=true,
  showstringspaces=false,
  frame=single,
  framerule=0.3pt,
  xleftmargin=0pt,
  xrightmargin=0pt,
  aboveskip=4pt,
  belowskip=6pt
}

\newif\ifanon
\anonfalse

\begin{document}

\title{Agentic Proof and Property-Based Testing via Property-Templates in Data-Intensive Computing}

\ifanon
\author{\IEEEauthorblockN{Anonymous Author(s)}}
\else
\author{
\IEEEauthorblockN{Seongmin Lee\textsuperscript{*}\thanks{\textsuperscript{*}Seongmin Lee and Yaoxuan Wu contributed equally to this work.}}
\IEEEauthorblockA{\textit{University of California, Los Angeles}\\
Los Angeles, CA, USA \\
seongminlee@g.ucla.edu}
\and
\IEEEauthorblockN{Yaoxuan Wu\textsuperscript{*}}
\IEEEauthorblockA{\textit{University of California, Los Angeles}\\
Los Angeles, CA, USA \\
thaddywu@cs.ucla.edu}
\and
\IEEEauthorblockN{Miryung Kim}
\IEEEauthorblockA{\textit{University of California, Los Angeles}\\
Los Angeles, CA, USA \\
miryung@cs.ucla.edu}
}
\fi

\maketitle

\begin{abstract}
As the cost of code generation becomes cheaper with AI, the new bottleneck in software engineering has shifted to intent specification and validation. Overcoming this durability crisis of AI-driven coding requires more than traditional fuzzing: each candidate property must be both \emph{proven} correct over a model of the system and \emph{shown to hold} on the real implementation, making formal proof and systematic property-based testing (PBT) complementary. However, validating properties this way at scale requires solving two core subproblems: (1) verifying that candidate properties are indeed correct, and (2) operationalizing PBT without AI hallucination. We hypothesize that recurring property patterns, cast as {\em property templates}\textemdash {\em abstract, parameterized forms with holes}\textemdash address both subproblems at once.

This paper investigates the value of recurring property patterns in a target system, Apache Spark. In data-intensive scalable computing systems, numerous correctness properties arise from the principles of data partition, computation decomposition, and data-flow computation. For instance, a classic recurring pattern is aggregation decomposition, where, for all data $D$ and workloads $Q$, a global function executed on the entire dataset relates to a local function followed by a recombiner. We design an agentic, dual-track validation framework that uses property templates to formally verify their correctness in the Lean 4 theorem prover, and instantiate PBT templates as concretized executable PBTs. Our evaluation shows that property templates increase agentic proof engineering success by up to $2.6\times$ (avg:\,$1.6\times$) and reduce proof hallucinations by $59\%$. Template-guided PBT synthesis reduces intent misalignments from 22 to 1 and cuts synthesis cost by up to $5.7\times$ (avg:\,$3.8\times$). Template-guided synthesis further exceeds a state-of-the-art Spark fuzzer and approaches unguided LLM-based PBT on code coverage. Finally, comparing the two tracks is informative: for example, when a proof succeeds yet a PBT finds a counterexample, the mismatch identifies a gap between the formal model and the implementation.
\end{abstract}

\section{Introduction}
\label{sec:introduction}

Establishing that a software system behaves as intended rests on two complementary validation techniques.
This behavior is captured as a \emph{specification}---a conjectured property the system should satisfy.
\emph{Formal proof}~\cite{demouraLeanTheoremProver2015} mathematically verifies it against a model of the system, while \emph{property-based testing} (PBT)~\cite{claessenQuickCheckLightweightTool2000,goldsteinPropertyBasedTestingPractice2024,10.1007/978-3-319-22102-1_22} exercises the real implementation on many generated inputs.
The two are complementary: a proof determines whether the property is in fact correct, while PBT tests whether the implementation actually obeys it.
Applying \emph{both} therefore yields far stronger evidence than either alone.

A real software system’s correctness hinges on complex, end-to-end properties of its behavior. For example, a query should return the same results whether or not the engine optimizes it, for every input dataset, rather than merely producing the correct output for a single function. Many such properties must be validated, including the correctness of each optimization, rewrite, and equivalence on which the engine relies.
Validating even one property through a machine-checked proof, an executable test, or both is laborious. Repeating this process by hand, property after property, does not scale.
Language models can now \emph{propose} candidate properties at scale~\cite{maazAgenticPropertyBasedTesting2025a,lahiriIntentFormalizationGrand2026b}, but this only sharpens the validation problem: a proposed property may be subtly false, and a generated test may silently check something weaker than intended.
The problem we address is how to \emph{validate} these properties at scale by confirming that each is correct and exercising it as intended.

Data-intensive systems such as Apache Spark~\cite{zahariaApacheSparkUnified2016} are built on a few underlying principles---partitioning data across a cluster, decomposing computation over the partitions, and composing operators into dataflow pipelines.
These principles make a system's correctness properties \emph{recur}: the same structural relationship reappears as a \emph{family} of closely related properties across the API.
\emph{Aggregation decomposition} is one such family. It relates a global aggregate to the recombination of per-partition aggregates, with one member for every aggregator: a global \texttt{sum} equals the sum of per-partition sums, a global \texttt{max} equals the maximum of per-partition maxima, and so on.
A second family, \emph{UDF rewrite}, captures a different relationship. It states that an opaque user-defined function is equivalent to a built-in expression that the engine can optimize. For instance, a Python UDF that uppercases a string is equivalent to the built-in \texttt{upper}, with one member for every replaceable UDF.
Each family is also vast and not uniformly true: aggregation decomposition, for instance, does not hold for the \texttt{mean}, since a global mean does \emph{not} equal the mean of per-partition means. Telling such false members apart is hard at scale, as the family admits $37{,}971$ type-consistent instantiations, many of which type-check yet are false.
Validating so many candidates, true and false alike, by hand is infeasible.

We therefore exploit this recurrence with two pieces: a \emph{property template} that codifies a family's shared structure once as a parameterized form with typed holes, and \emph{agentic} methods that turn it into a machine-checked proof and an executable test of each concrete member.
A single template takes two forms over the same property: a \emph{proof template}, a parameterized Lean~4 theorem that reduces an instance's proof to one local obligation which lifts to the full property by construction; and a \emph{PBT template}, a generator that turns the same instance into a diverse, executable test of the real system.
For each form, the agent fills only the template's property-specific holes---rather than re-deriving a proof or re-authoring a test from scratch---so that the cost of validation is amortized across the whole family.

We build property templates for four families of Spark properties whose violation silently corrupts results: aggregation decomposition, UDF rewrite, higher-order expression rewrite, and operator subsumption. We evaluate them on 100 candidate properties per family, for 400 properties in total, against template-free baselines on both tracks.
On the proof track, a pre-verified proof structure raises machine-checked synthesis successes by up to $2.6\times$ (avg:\,$1.6\times$) and reduces hallucinated proofs, which compile but prove nothing of interest, by $59\%$, at \emph{lower} cost.
On the PBT track, fixing the test architecture cuts intent misalignments in synthesized tests (from 22 to 1) and LLM synthesis cost by up to $5.7\times$ (avg:\,$3.8\times$).
Finally, the two tracks corroborate each other on 130 properties that earn both a proof and a passing test, and disagree informatively elsewhere: a passing PBT without a proof identifies properties that may benefit from broader formal modeling, whereas a PBT counterexample to a proven property exposes a mismatch between the Lean model and PySpark runtime semantics.

This paper makes the following contributions:
\begin{itemize}[leftmargin=*]
  \item We identify the recurring principles behind the correctness properties of data-intensive systems and abstract each resulting \emph{family} of properties into a \emph{property template}: a parameterized form with typed holes that captures the family's shared proof and shared test in a single artifact; in our evaluation, four such templates produced 136 successfully synthesized proofs and 387 faithful PBTs.
  \item To validate properties at scale, we develop two template-guided agentic methods: an agentic prover that synthesizes machine-checked Lean~4 proofs over a model of PySpark’s core API, and an agentic test synthesizer that produces executable PySpark tests.
  \item Narrowing the LLM to the template's holes is decisive: on the proof track, synthesis successes rise by up to $2.6\times$ and hallucinations fall by $59\%$; on the PBT track, intent misalignments fall from 22 to 1 at up to $5.7\times$ lower cost, compared with template-free synthesis of the same property.
  \item We cross-validate the two tracks against each other: their agreement gives the strongest evidence available, while their disagreement surfaces model-to-implementation gaps neither finds alone.
\end{itemize}

\section{Background and Motivation}
\label{sec:motivation}

\subsection{Validating a Specification: Proof and PBT}
\label{subsec:proof-pbt}

Two methods establish that a system satisfies a specification: \emph{formal proof} and \emph{property-based testing} (PBT). A proof reasons deductively over a \emph{model} of the system to establish a property for \emph{all} inputs; written in a proof assistant such as Lean, Rocq, or Isabelle, it is verified by the tool itself, yielding a \emph{machine-checked} guarantee~\cite{demouraLeanTheoremProver2015,leroyFormalVerificationRealistic2009,kleinSeL4FormalVerification2009a}. PBT instead checks a property on the \emph{real implementation}~\cite{claessenQuickCheckLightweightTool2000}: rather than a single input and its expected output, as in a conventional test, one writes a \emph{property} meant to hold for all inputs together with a \emph{generator} that produces many varied, well-formed inputs, and checks the property on each, shrinking any counterexample to a minimal case. Because a single model and its generators yield large volumes of tests, PBT can drive \emph{system-level} testing of complete implementations, not just unit checks~\cite{hughesExperiencesQuickCheck2016}.
This scale makes PBT a practical vehicle for \emph{specification validation}: AWS's Kiro turns a natural-language specification into executable properties checked against the (increasingly AI-written) code~\cite{kiroCorrectnessPropertybasedTests2025}.

\begin{figure}[t]
\centering
\colorlet{gold}{orange!72!yellow}
\resizebox{\columnwidth}{!}{%
\begin{tikzpicture}[
    font=\small,>=Latex,node distance=0pt,
    pbox/.style={draw=cAgg,line width=0.7pt,rounded corners=2.5pt,fill=cAgg!4,
                 text=cAgg,align=center,inner sep=2.5pt,
                 text width=42mm,minimum height=9mm},
    ttl/.style={font=\small\bfseries},
    yarr/.style={-{Latex[length=2.6mm,width=3.6mm]},line width=2.4pt,gold,
                 rounded corners=3pt},
    alab/.style={font=\footnotesize,text=black,align=center},
    vlab/.style={font=\footnotesize\bfseries}]

  \node[pbox] (gen) at (0,3.45) {Generate a potential property};

  \node[ttl] (lt) at (-2.7,2.1) {Proof-Engineering};
  \node[ttl] (rt) at ( 2.7,2.1) {Property-based Testing};
  \node[pbox] (pf)  at (-2.7,1.25) {Prove whether it is a correct property on a model};
  \node[pbox] (pbt) at ( 2.7,1.25) {Test whether it holds true on a real system};

  \draw[gold,line width=2.4pt,rounded corners=3pt] (gen.south) -- (0,2.8);
  \draw[yarr] (0,2.8) -| (-2.7,2.32);
  \draw[yarr] (0,2.8) -| ( 2.7,2.32);

  \draw[yarr] (1.5,0.2) -- (-1.5,0.2);
  \node[alab] at (0,0.46) {Serve as a conjecture worth proving};
  \node[vlab] at (-3.05,0.33) {Proof: ??};
  \node[vlab] at ( 3.05,0.33) {PBT: \cmark};

  \draw[yarr] (-1.5,-0.55) -- (1.5,-0.55);
  \node[alab] at (0,-0.29) {Must hold in a real system};
  \node[vlab] at (-3.05,-0.42) {Proof: \cmark};
  \node[vlab] at ( 3.05,-0.42) {PBT: ??};
\end{tikzpicture}}
\caption{Proof and property-based testing each supply the evidence the other
lacks when validating a property.}
\label{fig:symbiosis}
\vspace{-1em}
\end{figure}
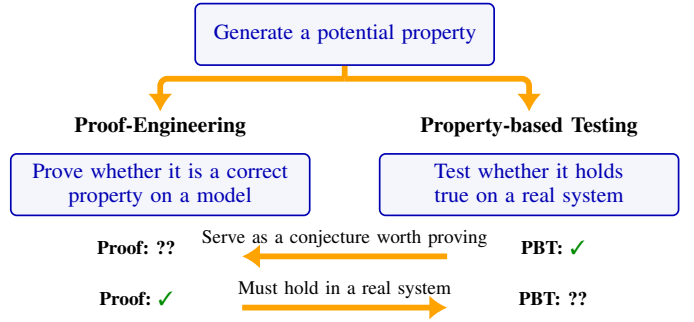

Proof and PBT are \emph{symbiotic} rather than redundant (Figure~\ref{fig:symbiosis}): a proof gives a \emph{sound guarantee} over a model of all inputs, while PBT gives \emph{empirical evidence} from the real system on many, so each reaches evidence the other cannot. A property that survives extensive testing is worth proving, both to extend the guarantee to all inputs and to delineate the scope the formal model must capture. A proved property, in turn, still calls for empirical evidence: should a test find a counterexample, either the proof is unsound or a gap separates the model from the implementation. Agreement along both tracks is the strongest evidence of all.

\subsection{Recurring Properties in DISC}

\paragraph{DISC: Data-Intensive Scalable Computing Systems}

\begin{figure*}[h]
\centering
\begin{tikzpicture}[font=\small,>=Latex,
  op/.style={draw=cAgg,fill=cAgg!75,text=white,font=\scriptsize,
             minimum width=6.5mm,minimum height=5.5mm,inner sep=1pt,rounded corners=1pt},
  dpart/.style={draw=cAgg,fill=cAgg!12,minimum width=4.2mm,minimum height=3.2mm,inner sep=0,
             path picture={\fill[cAgg!75] (path picture bounding box.north west)
                rectangle ([yshift=-0.8mm]path picture bounding box.north east);}},
  dres/.style={draw=cAgg,fill=cAgg!28,minimum width=4.2mm,minimum height=2.2mm,inner sep=0},
  shf/.style={draw,fill=red!8,rounded corners=2pt,font=\scriptsize,align=center},
  ttl/.style={font=\small\bfseries},
  nt/.style={font=\footnotesize,gray},
  flow/.style={->,line width=0.9pt},
  cdot/.style={->,densely dotted,line width=0.7pt,shorten >=1pt},
  sep/.style={gray!35}]

\node[ttl] at (2.3,3.72) {\textcircled{\footnotesize 1}\;\,DISC computation};
\node[nt] at (0.7,2.95) {DataFrame~$D$};
\node[nt] at (2.95,2.95) {workload~$Q$};
\draw[decorate,decoration={brace,amplitude=4pt},gray,line width=0.9pt]
  (0.42,2.58) -- (0.98,2.58);
\draw[decorate,decoration={brace,amplitude=4pt},gray,line width=0.9pt]
  (1.55,2.58) -- (4.35,2.58);
\node[dpart] (d1) at (0.7,2.2){};
\node[dpart] (d2) at (0.7,1.4){};
\node[dpart] (d3) at (0.7,0.6){};
\node[op] (mp)  at (2.0,2.0) {map};
\node[op] (scn) at (2.0,0.8) {scan};
\node[op] (jn)  at (3.0,1.4) {join};
\node[op] (ag)  at (3.95,1.4) {agg};
\draw[cdot] (d1.east) to[bend left=6] (mp.west);
\draw[cdot] (d2.east) to[bend left=2] (mp.west);
\draw[cdot] (d3.east) to[bend right=6] (scn.west);
\draw[flow] (mp)--(jn); \draw[flow] (scn)--(jn); \draw[flow] (jn)--(ag);

\node[ttl] at (8.3,3.72) {\textcircled{\footnotesize 2}\;\,The aggregation-decomposition family};
\begin{scope}[every node/.style={font=\footnotesize,inner sep=1pt}]
\node[anchor=east] (tl) at (7.0,2.5) {\aggfn{agg}(\dff)};
\node             (tm) at (7.2,2.5) {\relsym{R}};
\node[anchor=west] (tr) at (7.4,2.5)
     {\recfn{recombine}\textcolor{cRec}{(}\,\aggfn{agg}(\dfg)\textsf{,}\,\dots\textsf{,}\,\aggfn{agg}(\dfg)\,\textcolor{cRec}{)}};
\node[anchor=west,font=\footnotesize,text=gray] at (5.2,1.65) {global count $\geq$ largest per-partition count};
\node[anchor=east] at (7.0,1.2) {\aggfn{count}(\dff)};
\node             at (7.2,1.2) {\relsym{\geq}};
\node[anchor=west] at (7.4,1.2)
     {\recfn{max}\textcolor{cRec}{(}\,\aggfn{count}(\dfg)\textsf{,}\,\dots\textsf{,}\,\aggfn{count}(\dfg)\,\textcolor{cRec}{)}};
\node[anchor=west,font=\footnotesize,text=gray] at (5.2,0.5) {distinct values $=$ union of per-partition distinct sets};
\node[anchor=east] at (7.0,0.05) {\aggfn{unique}(\dff)};
\node             at (7.2,0.05) {\relsym{=}};
\node[anchor=west] at (7.4,0.05)
     {\recfn{$\cup$}\textcolor{cRec}{(}\,\aggfn{unique}(\dfg)\textsf{,}\,\dots\textsf{,}\,\aggfn{unique}(\dfg)\,\textcolor{cRec}{)}};
\end{scope}
\node[draw=cRec,line width=1pt,rounded corners=3pt,fit=(tl)(tm)(tr),inner sep=4pt] (box){};
\node[nt,cAgg,anchor=base] (cag) at (5.75,3.2) {aggregator};
\node[nt,cRel,anchor=base] (cre) at (7.2,3.2) {relation};
\node[nt,cRec,anchor=base] (crc) at (8.65,3.2) {recombiner};
\draw[cdot,cAgg] (5.75,3.05) -- (5.75,2.74);
\draw[cdot,cRel] (7.2,3.05) -- (7.2,2.74);
\draw[cdot,cRec] (8.55,3.05) -- (7.95,2.74);

\node[ttl] at (14.7,3.72) {\textcircled{\footnotesize 3}\;\,A cheaper, equivalent plan};
\node[dpart] (e1) at (12.85,3.13){};
\node[dpart] (e2) at (12.85,2.75){};
\node[dpart] (e3) at (12.85,2.37){};
\node[shf,minimum height=10mm,minimum width=6.5mm] (esh) at (14.05,2.75){shuffle};
\node[op] (eag) at (15.25,2.75){agg};
\draw[cdot] (e1.east) -- ([yshift=2mm]esh.west);
\draw[cdot] (e2.east) -- (esh.west);
\draw[cdot] (e3.east) -- ([yshift=-2mm]esh.west);
\draw[flow] (esh)--(eag);
\node[draw=cRec,rounded corners=3pt,line width=0.8pt,fit=(e1)(e3)(esh)(eag),inner sep=3pt] (ebox){};
\coordinate (etopc) at (16.32,0 |- ebox.north);
\node[cRec,font=\footnotesize,align=left,anchor=north west,text width=1.45cm,inner xsep=1pt,inner ysep=0pt] (esent) at ([xshift=-2pt,yshift=2pt]etopc)
  {expensive: all rows cross the shuffle};
\coordinate (ehy) at ([yshift=-3mm]ebox.north east);
\coordinate (evtop) at ([xshift=-1.5pt]esent.north west);
\coordinate (evbot) at ([xshift=-1.5pt]esent.south west);
\draw[cRec,line width=0.8pt] (evtop) -- (evbot);
\draw[cRec,line width=0.8pt] (ehy) -- (evtop |- ehy);
\node[dpart] (c1) at (12.85,0.7){};
\node[dpart] (c2) at (12.85,0.35){};
\node[dpart] (c3) at (12.85,0.0){};
\node[op,minimum width=5.5mm,minimum height=3mm] (ca1) at (13.75,0.7){agg};
\node[op,minimum width=5.5mm,minimum height=3mm] (ca2) at (13.75,0.35){agg};
\node[op,minimum width=5.5mm,minimum height=3mm] (ca3) at (13.75,0.0){agg};
\draw[cdot] (c1)--(ca1);
\draw[cdot] (c2)--(ca2);
\draw[cdot] (c3)--(ca3);
\node[shf,minimum height=9mm,minimum width=6.5mm] (csh) at (14.85,0.35){shuffle};
\draw[flow] (ca1.east) -- ([yshift=2mm]csh.west);
\draw[flow] (ca2.east) -- (csh.west);
\draw[flow] (ca3.east) -- ([yshift=-2mm]csh.west);
\node[op,minimum width=8.5mm] (crec) at (16.05,0.35){recomb};
\draw[flow] (csh)--(crec);
\node[draw=violet!75!black,rounded corners=3pt,line width=0.8pt,fit=(c1)(c3)(ca1)(ca3)(csh)(crec),inner sep=3pt] (cbox){};
\node[violet!75!black,font=\footnotesize,align=left,anchor=south west,inner ysep=1pt] (csent) at ([yshift=0.12cm]cbox.north west)
  {cheap: pre-aggregate;\\only partials cross the shuffle};
\coordinate (cvx) at ([xshift=8pt]csent.south west);
\draw[violet!75!black,line width=0.8pt] (csent.south east) -- (csent.south west);
\draw[violet!75!black,line width=0.8pt] (cvx) -- (cvx |- cbox.north);
\node[dres] (eo) at (16.0,1.85){};
\node[dres] (co) at (16.95,1.85){};
\node at (16.475,1.85) {\large$=$};
\draw[flow,rounded corners=2pt] (eag.east) -| (eo);
\draw[flow,rounded corners=2pt] (crec.east) -| (co);

\begin{scope}[every node/.style={font=\footnotesize,align=center,anchor=north,inner sep=1pt}]
\node[text width=5.3cm] at (2.1,-0.4)
  {A workload~$Q$ runs over a DataFrame~$D$ partitioned across a cluster as a lazy dataflow DAG of operators.};
\node[text width=7.4cm] at (8.475,-0.4)
  {Choosing an \textcolor{cAgg}{aggregator}, a \textcolor{cRel}{relation}, and a \textcolor{cRec}{recombiner} gives a member relating the global aggregation to recombining the partitions.};
\node[text width=4.6cm] at (14.6,-0.4)
  {When the \textcolor{cRel}{relation} is `$=$', the \mbox{engine} pre-aggregates below the shuffle---a cheaper, equivalent plan.};
\end{scope}

\draw[sep] (4.7,-1.45)--(4.7,3.55);
\draw[sep] (12.25,-1.45)--(12.25,3.55);
\end{tikzpicture}
\caption{An example recurring property family, \emph{aggregation decomposition},
shown end to end. \protect\dfg~denotes a DataFrame partition.}
\label{fig:aggdecomp}
\end{figure*}

Data-intensive scalable computing (DISC) systems such as Apache Spark~\cite{zahariaApacheSparkUnified2016} process datasets too large for a single machine by partitioning the data across a cluster and computing over the partitions in parallel. Unlike a classical relational database, where the user issues a declarative query and the engine owns the entire execution plan, a DISC program is an explicit pipeline of coarse-grained transformations---\texttt{map}, \texttt{filter}, \texttt{groupBy}, \texttt{join}, and user-defined functions---that the framework compiles into a distributed dataflow DAG and executes lazily (Figure~\ref{fig:aggdecomp}, left). Because this model interleaves relational operators with arbitrary user code and exposes partitioning to the program itself, the engine's freedom to optimize rests on structural invariants governing how operators commute, distribute, and recombine across partitions. 
Such invariants recur throughout DISC and hold for every input dataset and every upstream pipeline. Particularly important are equivalences that enable the system to replace an expensive computation with a cheaper one proved to produce the same result, such as by pre-aggregating below a shuffle.

\begin{table*}[h]
  \centering
  \footnotesize
  \caption{Recurring property families in data-intensive computing. Each family is a group of properties that share one structure; the \textbf{Description} highlights the components that vary across members (numbered and colored), and each \textbf{Example} gives members as tuples of those components.}
  \label{tab:templates}
  \begin{tabularx}{\textwidth}{@{}l !{\vrule} >{\hsize=1.02\hsize}X !{\vrule} >{\hsize=0.98\hsize\raggedright\arraybackslash}X@{}} \toprule
    \textbf{Family} & \textbf{Description} & \textbf{Example members} \\ \midrule
    \textsc{HOE}
      & For every input dataset and any surrounding operations, replacing \textcolor{blue!70!black}{[1.\,an expression]} with \textcolor{orange!80!black}{[2.\,a pointwise-equivalent rewrite]} leaves the DataFrame output unchanged, whatever operator consumes it.
      & \textcolor{blue!70!black}{[1.\,\texttt{size(reverse(arr))}]} $\equiv$ \textcolor{orange!80!black}{[2.\,\texttt{size(arr)}]} --- reversing preserves length.\par\smallskip
        \textcolor{blue!70!black}{[1.\,\texttt{array\_contains(arr,elem)}]} $\equiv$ \textcolor{orange!80!black}{[2.\,\texttt{array\_position(arr,}\allowbreak\texttt{elem)}\,$>$\,\texttt{0}]} --- present iff the position is positive. \\ \midrule
    \textsc{UDF}
      & For every input dataset and any surrounding operations, \textcolor{blue!70!black}{[1.\,a Python UDF expression]} and \textcolor{orange!80!black}{[2.\,an equivalent PySpark built-in]} are interchangeable in any DataFrame operation that takes a column expression.
      & \textcolor{blue!70!black}{[1.\,\texttt{abs(x)}]} $\equiv$ \textcolor{orange!80!black}{[2.\,\texttt{F.abs(x)}]} --- same absolute value.\par\smallskip
        \textcolor{blue!70!black}{[1.\,\texttt{x[i:j]}]} $\equiv$ \textcolor{orange!80!black}{[2.\,\texttt{F.substring(x,i+1,j-i)}]} --- same substring (1-based indices). \\ \midrule
    \textsc{AggDecomp}
      & For every input dataset and any preceding operations, \textcolor{blue!70!black}{[1.\,aggregating]} over the whole dataset stands in \textcolor{green!50!black}{[2.\,a fixed relation]} to \textcolor{orange!80!black}{[3.\,recombining]} the same aggregate computed per group.
      & global \textcolor{blue!70!black}{[1.\,\texttt{count}]} \textcolor{green!50!black}{[2.\,$\geq$]} \textcolor{orange!80!black}{[3.\,\texttt{max}]} of the per-group counts.\par\smallskip
        global \textcolor{blue!70!black}{[1.\,\texttt{unique}]} \textcolor{green!50!black}{[2.\,$=$]} \textcolor{orange!80!black}{[3.\,\texttt{union}]} of the per-group uniques --- distinct values. \\ \midrule
    \textsc{Subsump}
      & For every input dataset and any preceding operations, \textcolor{blue!70!black}{[1.\,an operator]}'s output stands in \textcolor{green!50!black}{[2.\,a fixed multiset relation]} to the rows it receives.
      & \textcolor{blue!70!black}{[1.\,\texttt{orderBy}]} output \textcolor{green!50!black}{[2.\,$=$]} input as multisets --- reorders only.\par\smallskip
        \textcolor{blue!70!black}{[1.\,\texttt{filter}]} output \textcolor{green!50!black}{[2.\,$\subseteq$]} input as multisets. \\ \bottomrule
  \end{tabularx}
\end{table*}

These invariants come in families. We identify four recurring families, with examples in Table~\ref{tab:templates}: \emph{higher-order expression rewrite} (\textsc{HOE}), where substituting a pointwise-equivalent expression preserves the output whatever operator consumes it; \emph{UDF rewrite} (\textsc{UDF}), where a Python user-defined function---opaque to the engine---rewrites to an equivalent built-in; \emph{aggregation decomposition} (\textsc{AggDecomp}), where an aggregate over the whole dataset equals recombining the same aggregate computed per partition; and \emph{operator subsumption} (\textsc{Subsump}), where an operator's output stands in a fixed multiset relation to its input. Figure~\ref{fig:aggdecomp} makes one family concrete: its center shows several members of aggregation decomposition, each licensing a cheaper plan.

A family's shared structure makes a system's true invariants easier to find: by varying its components, one generates candidate properties in bulk. \textsc{AggDecomp}, for instance, is realized by any two of PySpark's $117$ collection-reducing functions---an aggregator and a recombiner---related by any of $9$ comparison operators, so it alone yields $117 \times 117 \times 9 = 123{,}201$ candidates, $37{,}971$ of them type-consistent.

But this abundance is double-edged: the candidates are too many to check by hand, and sharing the structure is no guarantee of truth---many are false. In \textsc{AggDecomp}, a global \textcolor{blue!70!black}{\texttt{mean}} \textcolor{green!50!black}{$=$} the \textcolor{orange!80!black}{\texttt{mean}} of the per-group means silently fails whenever partitions differ in size; in \textsc{UDF}, Python's \textcolor{blue!70!black}{\texttt{x.strip()}} $\equiv$ \textcolor{orange!80!black}{\texttt{F.trim(x)}} silently fails on Unicode whitespace that Spark's ASCII-only \texttt{trim} ignores. Both candidate properties are well-typed, but both are false. 
Which candidates are genuine properties of the family must therefore be determined one at a time, because neither shared structure nor successful type-checking establishes that a candidate is true. Doing so at scale requires automated proof and testing.

\subsection{Pilot Study: Using LLMs for PBT Synthesis in DISC}
\label{subsec:pilot}

We first ask whether a SOTA LLM can synthesize systematic PBTs without DISC-specific structure. Using GPT-5.5 (medium reasoning), we generate $100$ PySpark PBTs sequentially, each via plan-then-code; the model sees prior properties' names and statements to avoid duplication but receives no templates, examples, or execution feedback.

The resulting tests are individually meaningful but collectively unsystematic in two ways. \emph{No workload variation}: no test varies the surrounding operator sequence or embeds the property in a generated upstream workload; only the input data changes. \emph{No UDF coverage}: no test creates or invokes a Spark UDF, so none relates a user-defined computation to a built-in. Property templates systematically target selected property forms, while basic generators vary the surrounding workloads; we further evaluate both in Section~\ref{sec:eval-pbt}.

\section{\tool: Property Templates for Dual-Track Validation}
\label{sec:approach}

We present \tool, a framework that validates a system's many correctness properties at scale by making both proving and property-based testing agentic. For each recurring \emph{property family} of Section~\ref{sec:motivation}---\textsc{HOE}, \textsc{UDF}, \textsc{AggDecomp}, and \textsc{Subsump} (Table~\ref{tab:templates})---\tool builds a \emph{property template}: it \emph{parameterizes} the family's shared structure---turning the components that vary across members into typed holes an agent fills---and pairs it with the machinery to prove and test each instance. Our key hypothesis is that such a template makes validating a family's many properties both accurate and cost-efficient: because the properties share a common proof structure and test-generation machinery, the agent has less to complete on its own, so each proof and test is cheaper to produce and easier to get right.

Each property template comes in two forms that share its holes---a \emph{proof template} for the proof track and a \emph{PBT template} for the PBT track. The proof template is a parameterized theorem that reduces each proof to a single local law, so the proof agent establishes the property over a model of PySpark's core API instead of reasoning about the whole computation from scratch. The PBT template is a generator interface that varies the surrounding workload and encodes the property check through holes, so the testing agent produces many tests that correctly operationalize the property on the real implementation. Filling a template's holes---for \textsc{AggDecomp}, the \textcolor{blue!70!black}{aggregator}, \textcolor{orange!80!black}{recombiner}, and \textcolor{green!50!black}{relation}---yields a concrete property such as $(\textcolor{blue!70!black}{\texttt{count}}, \textcolor{orange!80!black}{\texttt{max}}, \textcolor{green!50!black}{\geq})$ that both forms carry, validated along a \emph{proof track} (Section~\ref{sec:proof-synthesis}) and a \emph{PBT track} (Section~\ref{sec:pbt-synthesis}).

\vspace{.5em}
\noindent \emph{Generation of Candidate Properties.}
Before detailing the two tracks (Sections~\ref{sec:proof-synthesis} and~\ref{sec:pbt-synthesis}), we explain how the concrete candidate properties they validate are generated.
We prompt an LLM with each template's natural-language definition to generate candidate properties. For \textsc{AggDecomp}, the prompt reads:

\begin{list}{}{\setlength{\leftmargin}{1em}\setlength{\rightmargin}{1em}\setlength{\topsep}{2pt}}
\item[]\itshape \small
``You are a research engineer specialising in PySpark semantics. Propose one new aggregation-decomposability property following the triplet structure \texttt{global\_result \textcolor{green!50!black}{=} \textcolor{orange!80!black}{recombine}(\allowbreak groupBy(key).\allowbreak agg(\textcolor{blue!70!black}{fn}(val)))}.
Output the chosen \texttt{(\textcolor{blue!70!black}{aggregator}, \textcolor{orange!80!black}{recombine}, \textcolor{green!50!black}{relation})} ...''
\end{list}

\noindent and the LLM returns instances such as $(\textcolor{blue!70!black}{\texttt{sum}}, \textcolor{orange!80!black}{\texttt{sum}}, \textcolor{green!50!black}{=})$, $(\textcolor{blue!70!black}{\texttt{count}}, \textcolor{orange!80!black}{\texttt{sum}}, \textcolor{green!50!black}{=})$, and $(\textcolor{blue!70!black}{\texttt{count}}, \textcolor{orange!80!black}{\texttt{max}}, \textcolor{green!50!black}{\geq})$. A deduplication memory was used across iterations to keep the generated candidates diverse.

\subsection{Agentic Proof Synthesis with Proof Template}
\label{sec:proof-synthesis}

\vspace{.5em}
\noindent\textbf{Lean Set Up.}
Given a candidate property instantiated from a template, we synthesize its proof over a custom Lean~4 model of a subset of PySpark's core API. The model captures the core semantics a correctness property must refer to---how a DataFrame is represented, how a workload transforms it, and the structural laws that govern these transformations. In total it comprises 17 data types, 71 functions and relations, and 60 theorems; we detail its three layers, the domain-helper and lemma libraries, and representative Lean code below.

\begin{figure*}[t]
  \centering
  \includegraphics[width=\textwidth]{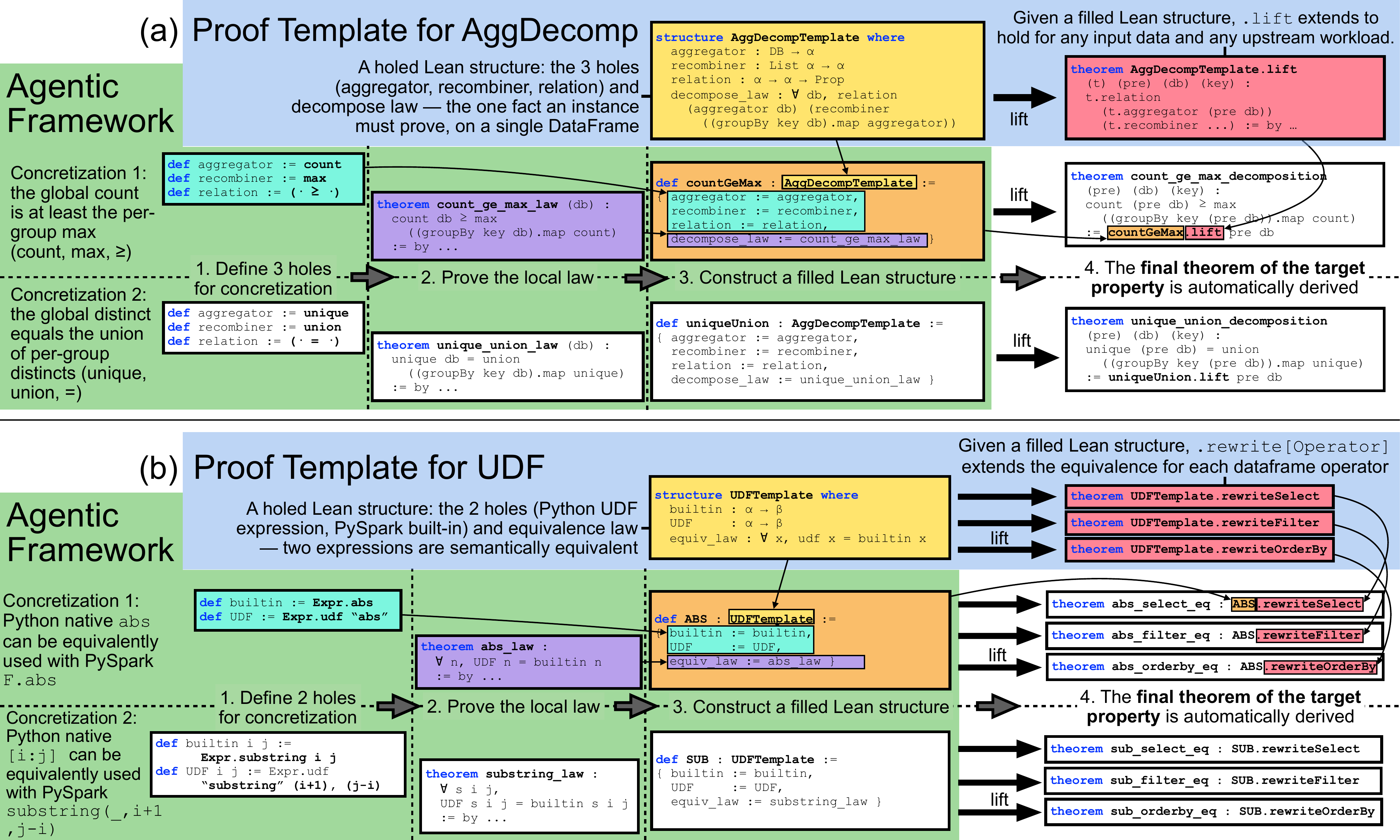}
  \caption{Proof templates and their agentic use for the two proof-track families,
    \textsc{AggDecomp}~(a) and \textsc{UDF}~(b), and two concretizations per template. The agent fills the holes and proves
    only the local law (steps~1--3); the template's pre-proved lift then yields the pipeline-level theorem automatically (step~4).}
  \label{fig:proof-templates}
  \vspace{-1em}
\end{figure*}

\vspace{.5em}
\noindent\textbf{Agentic Proof Synthesis.}
A property's proof is global---it must hold for every input dataset and every upstream workload---but it always splits into a \emph{local law}, an instance-specific fact about the components alone, and a \emph{lift} that carries the local law up to the global statement. We build each proof template around this split: we pre-prove the lift once per template and leave the \emph{local law} as the single obligation an instance must discharge, so the agent never re-derives the global argument. Figure~\ref{fig:proof-templates} shows this for \textsc{AggDecomp} and \textsc{UDF}: the proof template (above) is the holed structure---its holes, local law, and pre-proved lift. Given a concrete property, the agent takes four steps using the proof template: (1)~fills the template's holes; (2)~proves the matching local law against our Lean model, iterating on the compiler's feedback until it checks; (3)~bundles the components and proof into a filled template; and (4)~reads off the pipeline-level property from the pre-proved lift. What the local law asserts and what the lift ranges over differ by template, which we make concrete next.

\vspace{.5em}
\emph{Aggregation decomposition.}
The components are an aggregator, a recombiner, and a relation (Figure~\ref{fig:proof-templates}a). The local law \texttt{decompose\_law} states, on a single arbitrary DataFrame, that the global aggregation stands in the relation to recombining the per-group aggregations; a single pre-proved theorem \texttt{.lift} extends it to any input run through any upstream workload. For the running instance $(\texttt{count}, \texttt{max}, \geq)$, the agent proves only that a global count is at least the maximum of the per-group counts, and \texttt{.lift} yields the end-to-end property.

\vspace{.5em}
\emph{UDF rewrite.}
The components are a built-in expression and a user-defined function over the same columns (Figure~\ref{fig:proof-templates}b). The local law \texttt{equiv\_law} states that the two agree pointwise---on every input the UDF returns exactly what the built-in does. Here the lift is not a single theorem but a \emph{rewrite rule per DataFrame operator}: for each operator the expression may sit in (\texttt{select}, \texttt{filter}, \texttt{orderBy}, \dots), one pre-proved theorem licenses replacing the built-in with the UDF inside that operator without changing any surrounding pipeline's output. For an instance such as \texttt{abs} (Python \texttt{abs} replaced by PySpark \texttt{F.abs}), the agent proves only the pointwise equivalence, and every such rewrite follows from the template.

The two cases differ in where the proof effort falls. For \textsc{AggDecomp} the lift is small, so the local law the agent proves is the harder part; for \textsc{UDF} the pointwise equivalence is usually easy, and the labor shifts to the per-operator lift the template pre-proves. This pre-proved template, written once per property family and reused across its instances, averages 546 non-comment lines of Lean.

\vspace{.5em}

\newcommand{\lhole}[1]{{\setlength{\fboxsep}{0pt}\colorbox{yellow!20}{\strut\footnotesize\ttfamily #1}}}

\subsection{Agentic PBT Synthesis with PBT Template}
\label{sec:pbt-synthesis}

\begin{figure*}[t]
  \centering
  \begin{subfigure}[t]{0.49\textwidth}
    \centering
    \begin{lstlisting}[style=pyspark,basicstyle=\ttfamily\scriptsize,numbers=left,numberstyle=\tiny\color{gray},numbersep=4pt,xleftmargin=1.5em,aboveskip=0pt,belowskip=0pt,
      classoffset=1,morekeywords={AGG},keywordstyle=\color{blue!70!black}\bfseries,
      classoffset=2,morekeywords={RECOMBINE},keywordstyle=\color{orange!80!black}\bfseries,
      classoffset=3,morekeywords={COMPARE},keywordstyle=\color{green!50!black}\bfseries,
      classoffset=0,escapeinside={(*@}{@*)}]
import pyspark.sql.functions as F
class AggDecompTemplate:
  partition_col_type = (*@\lhole{DiscreteType}@*)  # hole: key type for groupBy (no float)
  agg_col_type       = (*@\lhole{NumericType}@*)  # hole: col type for aggregation
  def AGG(self, col):        # hole: aggregation
    return (*@\lhole{F.count(col)}@*)
  def RECOMBINE(self, col):  # hole: recombination
    return (*@\lhole{F.max(col)}@*)
  def COMPARE(self, a, b):   # hole: comparison
    return (*@\lhole{approx\_geq(a, b)}@*)
  def test(self):
    S = GenSchema(); D = GenData(S)
    W = GenWorkload(D)
    col_p = PickTypedCol(W, self.partition_col_type)
    col_a = PickTypedCol(W, self.agg_col_type)
    g = W.agg(self.AGG(col_a))
    l = (W.groupBy(col_p)
           .agg(self.AGG(col_a).alias(col_local))
           .agg(self.RECOMBINE(col_local)))
    assert self.COMPARE(g, l)
\end{lstlisting}
    \caption{\textsc{AggDecomp} template instantiated for the property that the global count is at least the maximum per-group count. The highlighted holes specify \texttt{F.count}, \texttt{F.max}, and the tolerance-aware comparison \texttt{approx\_geq}.}
    \label{lst:aggdecomp-template}
  \end{subfigure}\hfill
  \begin{subfigure}[t]{0.49\textwidth}
    \centering
    \begin{lstlisting}[style=pyspark,basicstyle=\ttfamily\scriptsize,numbers=left,numberstyle=\tiny\color{gray},numbersep=4pt,xleftmargin=1.5em,aboveskip=0pt,belowskip=0pt,
      classoffset=1,morekeywords={UDF,BUILTIN},keywordstyle=\color{blue!70!black}\bfseries,
      classoffset=0,escapeinside={(*@}{@*)}]
import pyspark.sql.functions as F
class UDFTemplate:
  input_col_types = (*@\lhole{[IntegerType]}@*)  # hole: col type(s) accepted by the UDF
  output_col_type = (*@\lhole{IntegerType}@*)    # hole: col type produced by the UDF
  def UDF(self, col):      # hole: user-defined function
    (*@\lhole{@udf(returnType=IntegerType())}@*)
    (*@\lhole{def fn(x): return \textasciitilde{}x}@*)
    return (*@\lhole{fn(col)}@*)
  def BUILTIN(self, col):  # hole: equivalent built-in
    return (*@\lhole{F.bitwiseNOT(col)}@*)
  def test(self):
    S = GenSchema(); D = GenData(S)
    W = GenWorkload(D)
    col_in = PickTypedCol(W, self.input_col_types)
    op  = GenOp(self.output_col_type)
    DW  = GenDownstream(self.output_col_type)
    r_u = DW(op(W, self.UDF(col_in)))
    r_b = DW(op(W, self.BUILTIN(col_in)))
    assert self.compare_DF(r_u, r_b)
\end{lstlisting}
    \caption{\textsc{UDF} template instantiated with a user-defined bitwise negation (\texttt{\textasciitilde{}x}) and its PySpark built-in counterpart, \texttt{F.bitwiseNOT}. The template evaluates both expressions within otherwise identical pipelines and compares their final DataFrames.}
    \label{lst:udf-template}
  \end{subfigure}
  \caption{PBT templates instantiated by the LLM. Highlighted regions are agent-filled property holes, while \texttt{Gen*} and \texttt{Pick*} denote reusable workload generators; the remaining test architecture is fixed by the template.}
  \label{fig:pbt-synthesis}
  \vspace{-1em}
\end{figure*}

\vspace{.5em}
\noindent\textbf{Agentic PBT Synthesis.}
A property-based test consists of two parts: a property-specific \emph{computation} and a reusable \emph{test architecture}. The architecture generates varied, well-formed workloads, executes both sides of the property, and compares their results end to end. Our PBT templates fix this architecture and expose only the property-specific computation as holes. The agent fills these holes, producing a test that runs directly against PySpark.
This separation addresses limitations of PBTs generated from scratch. An unaided LLM typically keeps the surrounding workload fixed and does not exercise UDFs (Section~\ref{subsec:pilot}); even when given a property, it may silently test a different claim (Section~\ref{sec:eval-pbt}). Figure~\ref{fig:pbt-synthesis} shows two templates with their property-specific holes filled and highlighted. We next describe the basic generators used to construct workloads, followed by the two templates.

\vspace{.5em}
\emph{Basic Generators.}
We build a library of basic generators for PySpark schemas, input data, DataFrame operators, and expression operations. These generators vary not only the input rows but also the workload surrounding the property under test. They compose operators and expressions into different sequences and pipeline shapes, allowing each PBT to exercise the property in a range of workload contexts.
The generators are also designed to produce well-formed workloads. They use schema-valid column references and type-compatible expressions, construct DataFrame dependencies as valid acyclic graphs, and enforce operator-specific constraints such as schema compatibility and join-key alignment.

\vspace{.5em}
\emph{Aggregation decomposition.}
The \textsc{AggDecomp} template (Figure~\ref{lst:aggdecomp-template}) exposes three holes, \textcolor{blue!70!black}{\textbf{AGG}}, \textcolor{orange!80!black}{\textbf{RECOMBINE}}, and \textcolor{green!50!black}{\textbf{COMPARE}}, along with two column-type declarations that select the grouping and aggregation columns from the generated workload \texttt{W}. The template constructs both the global and per-group computations from the same \texttt{W} and the same columns, leaving only the property-specific operations to the agent. As highlighted, the agent sets \texttt{AGG} to \texttt{F.count}, \texttt{RECOMBINE} to \texttt{F.max}, and \texttt{COMPARE} to \texttt{approx\_geq}.

\vspace{.5em}
\emph{UDF rewriting.}
The \textsc{UDF} template (Figure~\ref{lst:udf-template}) exposes two holes, \textcolor{blue!70!black}{\textbf{UDF}}, the function under test, and \textcolor{blue!70!black}{\textbf{BUILTIN}}, its claimed built-in equivalent, along with input and output column types. The output type also determines where the expression can appear; for example, a Boolean output allows operators such as \texttt{filter}. The template samples an upstream workload \texttt{W}, a compatible operator \texttt{op}, and a downstream workload \texttt{DW}. It then runs two branches that share the same workloads and operator and differ only in whether they use \texttt{UDF} or \texttt{BUILTIN}. As highlighted, the agent fills \texttt{UDF} with a \texttt{@udf}-decorated bitwise negation and \texttt{BUILTIN} with \texttt{F.bitwiseNOT}.

Our PBT-track implementation includes 7,899 LOC of generator code and 1,109 LOC of template code. The generators support 12 column types, including arrays and maps, 23 DataFrame operators, and 93 expression operations spanning aggregation, string, array, window, and higher-order operations.

\section{Evaluation}
\label{sec:evaluation}

We evaluate our framework across four property families over PySpark: \textsc{HOE}, \textsc{UDF}, \textsc{AggDecomp}, and \textsc{Subsump}. For each, the generation procedure of Section~\ref{sec:approach} instantiates 100 candidate properties---400 in total---the fixed population every experiment below runs on, presented identically to each configuration. Produced automatically from the template's definition rather than by hand, they span the family broadly and keep the hard and/or invalid properties, so that the evaluation reflects the real-world challenge of property validation.

\subsection{Proof Synthesis}
\label{subsec:eval-proof}

\paragraph{Research Questions}
We ask two questions about the effect of the property template on
LLM-driven proof synthesis:
\begin{itemize}
  \item[\textbf{RQ1.}] Does the property template improve proof synthesis
    efficiency, measured in both success rate and LLM cost per property attempted?
  \item[\textbf{RQ2.}] Does the property template reduce proof hallucinations?
\end{itemize}

A successful \texttt{lake build} certifies the proof, not the definition it rests on: it guarantees the theorem follows from its definitions, not that they encode the intended property. An agent can thus pass the compiler while mis-stating the property, assuming vacuous hypotheses, or smuggling in an unsound shortcut---a compiling but misdirected proof we term a \emph{proof hallucination}. As the compiler cannot catch this, we manually inspect every compiling proof.

\begin{table}[t]
  \centering
  \caption{Proof synthesis outcomes per property family, over the in-scope properties of each. \emph{Compiles} counts proofs that pass \texttt{lake build}; \emph{Success} counts proofs that additionally pass a manual inspection of the generated Lean file for cheat patterns (trivial relation, input collapse, etc.); \emph{Hallucinated} is the remainder. Cost is reported per property attempted.}
  \label{tab:proof-results}
  \setlength{\tabcolsep}{3pt}
  \resizebox{\columnwidth}{!}{%
  \begin{tabular}{llrrrr}
    \toprule
    \makecell[l]{Property\\Family (\# props)} & Config & Compiles & Success & Hallucinated & \makecell{Cost/prop\\(USD)} \\
    \midrule
    \multirow{2}{*}{\textsc{HOE} (37)}
      & \textsc{Template} & 28 & 28~(100\%) & 0~(\phantom{0}0\%) & \$0.78 \\
      & \textsc{NoTemp} & 17 & 17~(100\%) & 0~(\phantom{0}0\%) & \$1.02 \\
    \midrule
    \multirow{2}{*}{\textsc{UDF} (68)}
      & \textsc{Template} & 54 & 54~(100\%) & 0~(\phantom{0}0\%) & \$0.77 \\
      & \textsc{NoTemp} & 28 & 21~(\phantom{0}75\%) & 7~(25\%) & \$1.02 \\
    \midrule
    \multirow{2}{*}{\textsc{AggDecomp} (89)}
      & \textsc{Template} & 13 & 6~(\phantom{0}46\%) & 7~(54\%) & \$1.03 \\
      & \textsc{NoTemp} & 15 & 5~(\phantom{0}33\%) & 10~(67\%) & \$1.17 \\
    \midrule
    \multirow{2}{*}{\textsc{Subsump} (49)}
      & \textsc{Template} & 48 & 48~(100\%) & 0~(\phantom{0}0\%) & \$0.42 \\
      & \textsc{NoTemp} & 47 & 47~(100\%) & 0~(\phantom{0}0\%) & \$0.48 \\
    \bottomrule
  \end{tabular}}
  \vspace{-1em}
\end{table}

\paragraph{Experimental Setup}
Of the 100 properties per family, we retain those that fall within the scope of our Lean~4 model of PySpark's core API (Section~\ref{sec:proof-synthesis}; Table~\ref{tab:proof-results}). Excluded candidates leave the scope for three reasons: (i)~primitives absent from the \texttt{Expr} model (higher-order array ops, IEEE-754/NaN, regex); (ii)~aggregators with no sound pure-functional model (\texttt{first}/\texttt{last}, \texttt{try\_sum}); or (iii)~operations outside the single-input pipeline (two-input joins), leaving 243 of 400 candidates in scope.

We compare two configurations of the same proof-synthesis agent: an LLM that explores the Lean formalization through the lean-lsp MCP toolset and iteratively drafts and revises Lean code until it obtains a complete, machine-checked proof of the target property. The two configurations differ in a single respect: whether the agent is given the relevant property template (Section~\ref{sec:proof-synthesis}). \textsc{Template} hands the agent the matching property template and instructs it to instantiate and apply it. \textsc{NoTemp} withholds the template---it is neither available to nor mentioned to the agent---so the agent must produce the whole proof on its own. In both configurations the agent runs under the same budget (GPT-5.5, medium reasoning, max output tokens=65,536, up to 24 agentic turns) and is asked to prove the same property. The turn cap is raised to 32 for \textsc{AggDecomp}, whose proofs do not appear within 24 turns.

We report three metrics. \emph{Compiles} is the number of properties whose final \texttt{.lean} file passes \texttt{lake build} with no \texttt{sorry}. \emph{Success} (a \emph{synthesis success}) is the subset of compiling proofs that we additionally judge, by manual inspection of the generated Lean file, to actually establish the property's intended claim rather than a trivially-true substitute; \emph{Hallucinated} is the remaining compiling proofs. \emph{Cost} is the client-side USD spent on LLM calls at GPT-5.5 rates (\$5/M input, \$30/M output, reasoning included in output), reported per property attempted. Because both configurations are evaluated on the same in-scope properties, we test their difference in \emph{Success} with a paired McNemar test~\cite{mcnemarNoteSamplingError1947}, both overall and per family.

\paragraph{Results}

\begin{figure}[t]
  \vspace{-1em}
  \centering
  \begin{subfigure}[b]{0.22\columnwidth}
    \centering
    \includegraphics[width=\linewidth]{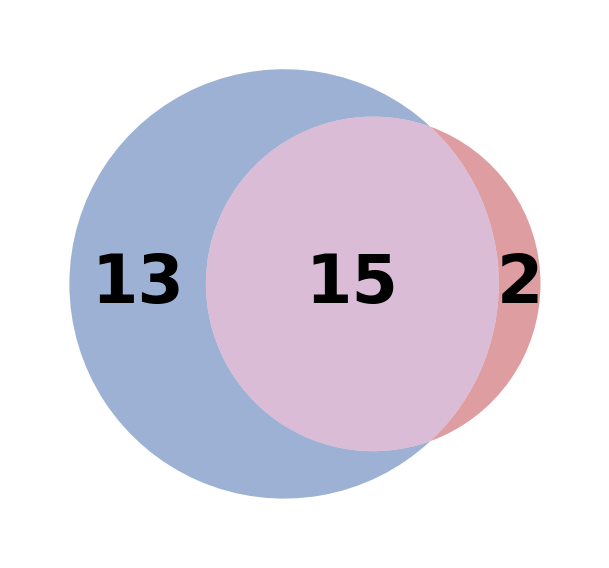}
    \caption{\textsc{HOE}}
    \label{fig:venn-hoe}
  \end{subfigure}\hfill
  \begin{subfigure}[b]{0.22\columnwidth}
    \centering
    \includegraphics[width=\linewidth]{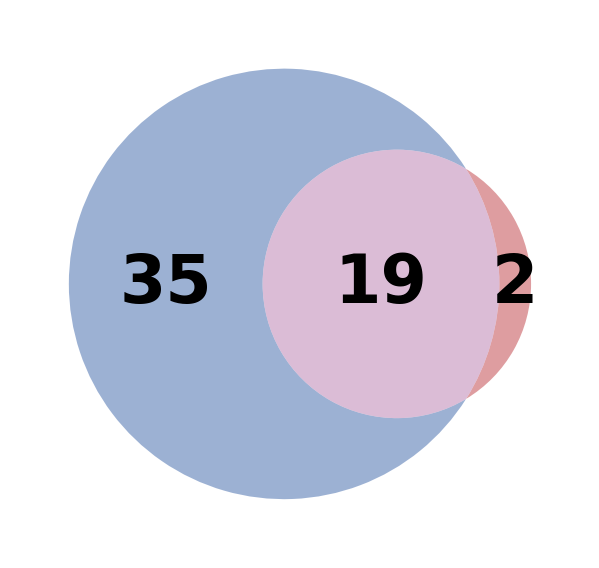}
    \caption{\textsc{UDF}}
    \label{fig:venn-udf}
  \end{subfigure}\hfill
  \begin{subfigure}[b]{0.30\columnwidth}
    \centering
    \includegraphics[width=\linewidth]{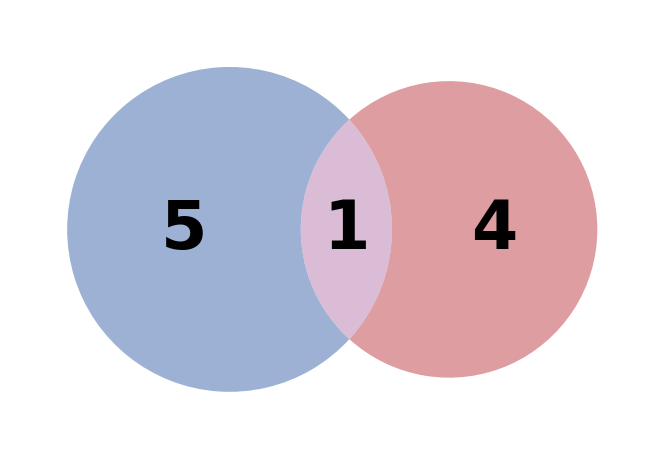}
    \caption{\textsc{AggDecomp}}
    \label{fig:venn-aggdecomp}
  \end{subfigure}\hfill
  \begin{subfigure}[b]{0.22\columnwidth}
    \centering
    \includegraphics[width=\linewidth]{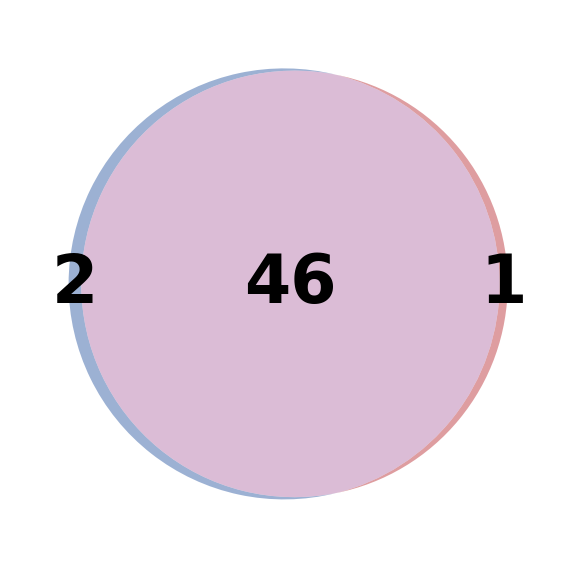}
    \caption{\textsc{Subsump}}
    \label{fig:venn-subsump}
  \end{subfigure}
  \caption{Synthesis successes per property family: by \textsc{Template} only (blue), \textsc{NoTemp} only (red), or both (centre).}
  \label{fig:proof-venn}
  \vspace{-1em}
\end{figure}

Across the four property families, \textsc{Template} produces 136 synthesis successes against \textsc{NoTemp}'s 90 (Table~\ref{tab:proof-results}), a per-family improvement of $1.0$--$2.6\times$ (averaging $1.6\times$). \textsc{Template} is also cheaper per property attempted in every family, with \textsc{NoTemp} spending $14$--$32\%$ more (averaging $23\%$). Hallucinations appear in only two families: \textsc{Template} hallucinates only on \textsc{AggDecomp}, where its rate stays below \textsc{NoTemp}'s ($54\%$ vs.\ $67\%$ of compiling proofs), while \textsc{NoTemp} also hallucinates on \textsc{UDF}; on the remaining two families neither configuration hallucinates at all. Figure~\ref{fig:proof-venn} shows the property-level overlap: 55 properties are successfully synthesized by \textsc{Template} alone, 81 by both configurations, and just 9 by \textsc{NoTemp} alone. By the paired McNemar test, this overall advantage is significant ($p<10^{-6}$). \textsc{Template} therefore covers all but 9 of the 145 properties that either configuration can synthesize successfully.

\paragraph{Per property family analysis}
These numbers point to \emph{where the LLM's effort goes}. The template's benefit is large for \textsc{HOE} and \textsc{UDF}---synthesis successes rise $1.6\times$ and $2.6\times$, and \textsc{NoTemp} pays about a third more per property---but small for \textsc{AggDecomp} and \textsc{Subsump}, where the two configurations land within a single synthesis success of each other and only the cost gap remains; the paired McNemar test accordingly finds the success difference significant for \textsc{HOE} ($p<0.01$) and \textsc{UDF} ($p<10^{-6}$) but not for \textsc{AggDecomp} or \textsc{Subsump}. The split reflects how much of the proof the template takes over. For \textsc{HOE} and \textsc{UDF} the per-property sub-goal is the easy part; the real work is lifting it to the operation level and proving it across the many DataFrame operators a pipeline may apply, and that lifting is exactly what the pre-verified template supplies (Section~\ref{sec:proof-synthesis}). The agent is thus spared designing the overall proof strategy---the step where LLM proof synthesis most often slips into unstructured low-level tactic search~\cite{jiangDraftSketchProve2023}. \textsc{NoTemp}, given no template, must invent both the high-level structure and its Lean proof within one turn budget, which both narrows its coverage and raises its cost. For \textsc{AggDecomp} and \textsc{Subsump} the sub-goal already operates over the DataFrame itself, so little is left to lift---only the extension to an arbitrary initial input and prefix workload---and \textsc{Template} and \textsc{NoTemp} come out nearly even on synthesis successes, though \textsc{Template} stays cheaper. The two differ in absolute difficulty, though: \textsc{Subsump} is easy enough that both nearly always succeed, while \textsc{AggDecomp}'s multi-input aggregation reasoning is challenging enough that both configurations fail more often.

\vspace{.5em}
\simplebox{black!5}{%
  \noindent\textbf{Answer to RQ1.} The property template raises synthesis successes from 90 to 136---up to $2.6\times$ per family, averaging
  $1.6\times$. \textsc{Template} is also cheaper per property in every
  family, with \textsc{NoTemp} spending $14$--$32\%$ more (avg: $23\%$).
}
\vspace{.5em}

\paragraph{Hallucination analysis}

Hallucination surfaces only in \textsc{UDF} and \textsc{AggDecomp}, and for a common reason. It arises from slack in translating a property's natural-language description into Lean: when each element of the description maps one-to-one onto a single definition in the Lean model---as it does for \textsc{HOE} and \textsc{Subsump}---there is little room for a plausible-looking but incorrect encoding. \textsc{UDF}, which must encode an arbitrary lambda expression, and \textsc{AggDecomp}, which composes more complex multi-input computations, leave much more room. The template removes the resulting hallucination for \textsc{UDF} but leaves it for \textsc{AggDecomp}; we examine each in turn.

For \textsc{UDF}, \textsc{NoTemp} frequently hallucinates through \emph{carrier collapse}: it reduces a multi-column operation to a single input or a constant so that the two sides coincide trivially---modelling 2-argument null-safe equality as \texttt{x == x}, a three-column \texttt{array\_join} as \texttt{"x|x|x"}, or a four-column except-cardinality as \texttt{size(except([x,x],[x,x])) = 0}. 7 of \textsc{NoTemp}'s 28 compiling \textsc{UDF} proofs ($25\%$) collapse this way. In contrast, \textsc{Template} produces none: all 54 of its compiling proofs are synthesis successes. The template hands the agent the proof's high-level structure outright, so its effort goes to a narrower sub-problem instead of the low-level tactic search that \textsc{NoTemp} often falls into, and trivial shortcuts like carrier collapse stop being a viable strategy.

Unlike \textsc{UDF}, for \textsc{AggDecomp} both configurations hallucinate, and in the same two forms: a \emph{tautological relation} ($P \vee \neg P$, a full trichotomy, \texttt{x = x}, a vacuous implication) and a \emph{degenerate aggregator} (a constant in place of the data-dependent statistic, a different statistic, or the wrong null/defined gate). 7 of \textsc{Template}'s 13 compiling proofs and 10 of \textsc{NoTemp}'s 15 hallucinate---\textsc{Template}'s almost all tautological relations, \textsc{NoTemp}'s spread across tautological relations and, more often, constant aggregators. Here the template does not separate the two configurations: its \texttt{DecompTriple} hands the relation and the aggregator to the agent to define---which \textsc{UDF}'s template does not---so even with the template in place the agent retains many angles from which to cheat. \textsc{NoTemp} hallucinates somewhat more ($67\%$ vs.\ $54\%$ of compiles), but neither is clean.

\vspace{.5em}
\simplebox{black!5}{%
  \noindent\textbf{Answer to RQ2.} The property template sharply reduces
  hallucination: it eliminates \textsc{NoTemp}'s \textsc{UDF} hallucinations (7
  proofs to none) and cuts \textsc{AggDecomp}'s (10 to 7), lowering the total
  across the families from 17 (\textsc{NoTemp}) to 7 (\textsc{Template})---a $2.4\times$
  reduction.
}

\subsection{PBT Synthesis}
\label{sec:eval-pbt}

\paragraph{Research Questions}
\begin{itemize}
  \item[\textbf{RQ3.}] On the same concrete property, does the property template improve PBT synthesis accuracy and reduce LLM cost, compared with synthesizing the test without it?
  \item[\textbf{RQ4.}] Does confining synthesis to template families sacrifice the behavioral coverage against unrestricted DISC testing\textemdash fuzzing and unguided, LLM-generated PBT?
\end{itemize}

As in \S\ref{subsec:eval-proof}, RQ3 isolates the template's effect---the same property synthesized with and without it. RQ4 instead probes a potential cost of the approach: comparing template-guided synthesis against two unrestricted baselines, it asks whether confining generation to a few predefined families sacrifices the behavioral coverage they attain.

\begin{table}[t]
  \centering
  \caption{PBT synthesis outcomes per property family, one attempt per property. A test is \emph{Faithful} if it executes and matches its natural-language description; otherwise it fails as \emph{Non-exec.}\ (does not execute) or \emph{NL-mis.}\ (executes but diverges from it).}
  \label{tab:pbt-instantiation-results}
  \footnotesize
  \setlength{\tabcolsep}{3pt}
  \vspace{-0.4em}
  \resizebox{\columnwidth}{!}{%
  \begin{tabular}{llrrrr}
    \toprule
    \makecell[l]{Property\\Family (\# props)}
      & Config.
      & Non-exec.
      & NL-mis.
      & Faithful
      & \makecell{Cost/prop\\(USD)} \\
    \midrule
    \multirow{3}{*}{\textsc{HOE} (100)}
      & \textsc{Template} &  3 & 0 & 97 (97.0\%) & \$0.031 \\
      & \textsc{GenOnly}  & 10 & 3 & 87 (87.0\%) & \$0.122 \\
      & \textsc{NoTemp}   &  5 & 3 & 92 (92.0\%) & \$0.172 \\
    \midrule
    \multirow{3}{*}{\textsc{UDF} (100)}
      & \textsc{Template} &  2 &  0 & 98 (98.0\%) & \$0.034 \\
      & \textsc{GenOnly}  & 13 & 15 & 72 (72.0\%) & \$0.158 \\
      & \textsc{NoTemp}   &  2 & 16 & 82 (82.0\%) & \$0.194 \\
    \midrule
    \multirow{3}{*}{\textsc{AggDecomp} (100)}
      & \textsc{Template} &  2 & 0 &  98 (98.0\%)  & \$0.060 \\
      & \textsc{GenOnly}  & 16 & 1 &  83 (83.0\%)  & \$0.124 \\
      & \textsc{NoTemp}   &  0 & 0 & 100 (100.0\%) & \$0.152 \\
    \midrule
    \multirow{3}{*}{\textsc{Subsump} (100)}
      & \textsc{Template} & 5 & 1 & 94 (94.0\%) & \$0.151 \\
      & \textsc{GenOnly}  & 9 & 3 & 88 (88.0\%) & \$0.167 \\
      & \textsc{NoTemp}   & 0 & 3 & 97 (97.0\%) & \$0.195 \\
    \bottomrule
  \end{tabular}}
  \vspace{-1em}
\end{table}

\paragraph{Experimental Setup (RQ3)}
For RQ3, we evaluate all 400 properties across three configurations, all single-round with GPT-5.5 (medium reasoning) (Table~\ref{tab:pbt-instantiation-results}). A PBT template supplies two reusable parts---workload \emph{generators} and a \emph{harness} that wires them around the property (\S\ref{sec:pbt-synthesis}). \textsc{Template} supplies both, leaving the LLM only the property holes; \textsc{GenOnly} supplies only the generators; \textsc{NoTemp} neither. A PBT is \emph{faithful} only if it both executes and matches the property's natural-language description; otherwise it fails as \emph{non-executable} or \emph{NL-misaligned} (executes but diverges).

\paragraph{RQ3 Faithfulness Results}
\textsc{Template} is the most faithful configuration---387/400 (96.8\%), against \textsc{NoTemp}'s 371/400 (92.8\%) and \textsc{GenOnly}'s 330/400 (82.5\%) (Table~\ref{tab:pbt-instantiation-results}). NL misalignments drive the gap: \textsc{Template} produces just 1, versus 22 for \textsc{NoTemp} and 22 for \textsc{GenOnly}. \textsc{NoTemp}'s 22 cluster where a property's semantics are subtle---19 boundary-semantics errors (16 in \textsc{UDF}, where the UDF and built-in diverge on null, NaN, or encoding; 3 in \textsc{HOE}, on null handling across the two sides) and 3 in \textsc{Subsump} that substitute row count for multiset containment. By fixing the harness structure, the template closes this room for misencoding.

Generator access alone does not help: \textsc{GenOnly} (82.5\%) falls below even \textsc{NoTemp} (92.8\%), as assembling a complete PBT from building blocks yields far more non-executable failures (48 vs.\ 7). The template does not eliminate such non-executable failures either---they stem from hallucinated Python and PySpark API calls, a code-level issue independent of the PBT structure.

\paragraph{RQ3 Token Cost Results}
\textsc{Template} also cuts cost, most where it constrains synthesis most tightly: per attempt it is 5.7$\times$ cheaper on \textsc{UDF} and 5.5$\times$ on \textsc{HOE} (\$0.03 vs.\ \$0.17--\$0.19), tapering to 2.5$\times$ on \textsc{AggDecomp} and 1.3$\times$ on \textsc{Subsump}. Most of the saving is in reasoning tokens---a 2.5$\times$ drop, from 4{,}500 to 1{,}800. \textsc{GenOnly}, still assembling the full harness, achieves only a 1.1–1.5$\times$ cost reduction.

\vspace{.5em}
\simplebox{black!5}{%
  \noindent\textbf{Answer to RQ3.}
  \textsc{Template} increases faithful PBT synthesis from 92.8\% to 96.8\%, reducing NL misalignments from 22 to 1 and cutting cost by up to 5.7$\times$ for \textsc{UDF} and \textsc{HOE}.
}

\paragraph{Experimental Setup (RQ4)}
The two unrestricted baselines are \textsc{LLM-PBT} and \textsc{CometFuzz}~\cite{cometfuzz_github}. \textsc{LLM-PBT} is the unconstrained LLM generation of our pilot study (Section~\ref{subsec:pilot}): GPT-5.5 synthesizes 100 PySpark PBTs with no template or pre-specified family (full prompt in supplemental). \textsc{CometFuzz} is a Spark fuzzer; it asserts no properties, so we compare it on code coverage alone. We assess three coverage dimensions: (i)~API coverage of PySpark expression operations and DataFrame operators (vs.\ \textsc{LLM-PBT}); (ii)~code coverage of Spark's \texttt{catalyst} and \texttt{execution} modules (vs.\ both); and (iii)~overlap in the kinds of properties tested (vs.\ \textsc{LLM-PBT}).

\begin{figure}[t]
  \centering
  \includegraphics[width=\linewidth]{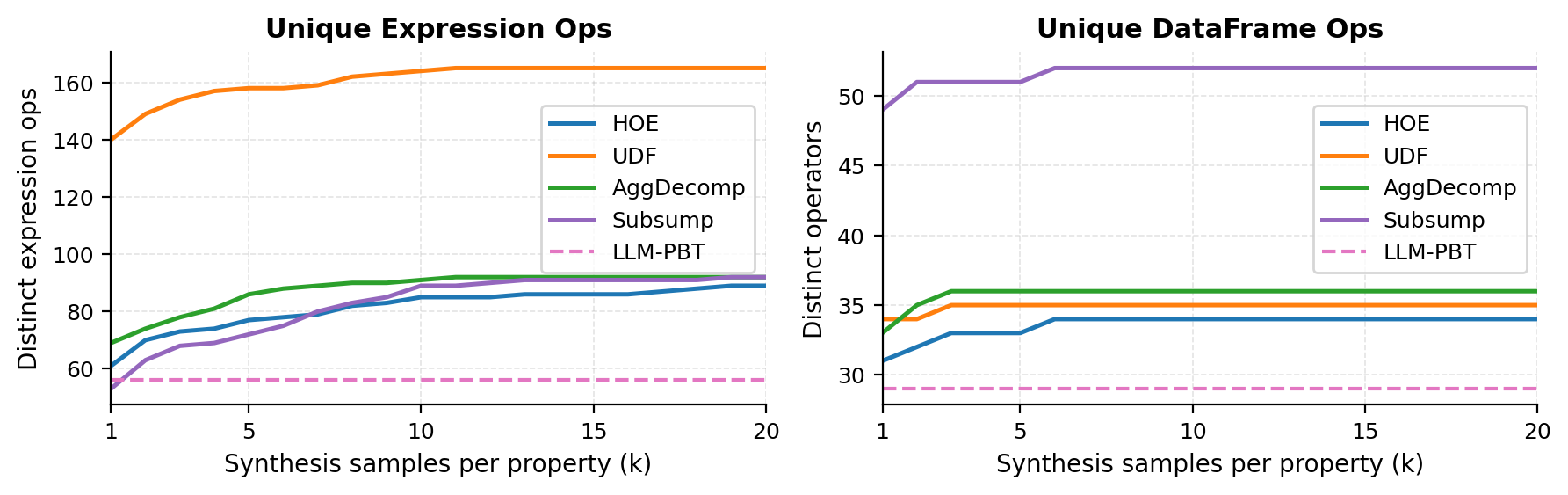}
  \caption{API coverage over the first 100 properties per family as a function of executions per property~($k$): unique expression operations (left) and DataFrame operators (right). Template-guided curves grow with~$k$; \textsc{LLM-PBT} (dashed) is flat.}
  \label{fig:pbt-api-coverage}
\end{figure}
\begin{figure}[t]
  \centering
  \includegraphics[width=\linewidth]{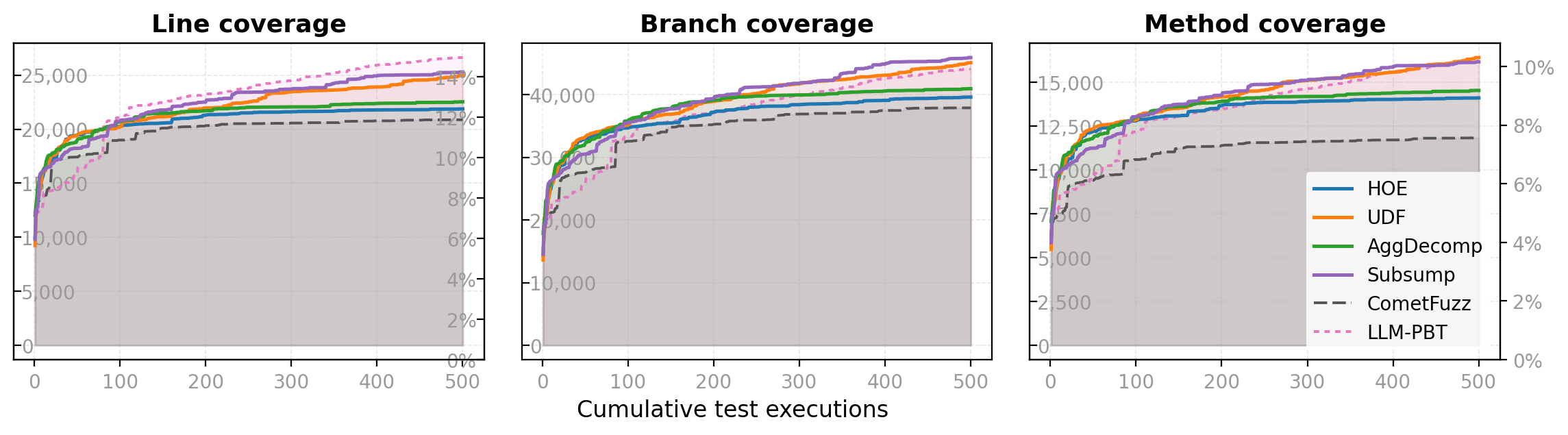}
  \vspace{-2ex}
  \caption{Line, branch, and method code coverage of Spark's \texttt{catalyst} and \texttt{execution} modules as a function of cumulative test executions. Template families and \textsc{LLM-PBT}: 100~PBTs $\times$ 5~executions; \textsc{CometFuzz}: 500~fuzz iterations.}
  \label{fig:pbt-coverage}
  \vspace{-1ex}
\end{figure}

\paragraph{API coverage}
\label{sec:rq4-results}
All four template families exercise more PySpark expression operations than \textsc{LLM-PBT}'s 56, ranging from 85 to 164 (Figure~\ref{fig:pbt-api-coverage}); \textsc{UDF} leads at 164, as each property pairs a UDF with a specific built-in. For DataFrame operators, \textsc{Subsump} reaches 52 against \textsc{LLM-PBT}'s 36, while the other three are comparable (34--36). Coverage grows with the number of executions per property ($k$) because each template execution samples a fresh workload, whereas \textsc{LLM-PBT}'s fixed workload does not vary with $k$.

\paragraph{Code coverage}
On Spark's \texttt{catalyst} and \texttt{execution} modules (Figure~\ref{fig:pbt-coverage}), template-guided synthesis \emph{exceeds} the \textsc{CometFuzz} fuzzer in every family, on both line coverage (12.5--14.4\% vs.\ 11.9\%) and method coverage (9.0--10.5\% vs.\ 7.5\%). Against \textsc{LLM-PBT} (15.2\% line, 10.3\% method) the families split: \textsc{UDF} and \textsc{Subsump} come within a percentage point on both and edge ahead on branch coverage (2--4\% higher), whereas \textsc{HOE} and \textsc{AggDecomp} trail by 2.3--2.7 points on line and 1.1--1.3 on method. These gains come from the generators, which vary input schemas and surrounding operations to reach broad engine paths within each family. Unlike a fuzzer, these tests also assert the property they target.

\paragraph{Property-space overlap}
Of the 100 \textsc{LLM-PBT} properties, 32 broadly align with our families (12 HOE, 7 AggDecomp, 13 Subsump, 0 UDF); the remaining 68 are single-operation expected-output (55), operator-algebra (5), round-trip (6), and execution-invariance (2) properties (full breakdown in supplemental). This reveals a breadth–depth tradeoff: direct synthesis spans more property forms, while templates instantiate reusable families systematically. The contrast is clearest for UDFs: none of the 100 \textsc{LLM-PBT} tests uses a Spark UDF, whereas the UDF template yields 98 faithful UDF-to-builtin PBTs.

\vspace{.5em}
\simplebox{black!5}{%
\noindent\textbf{Answer to RQ4.}
Template-guided synthesis exceeds \textsc{CometFuzz} on every coverage metric and approaches \textsc{LLM-PBT} (up to 4\% higher on branch), while exclusively covering UDF-to-builtin correspondence---a class unguided synthesis never generates.
}

\subsection{Cross-Validation: PBT Against Formal Proofs}
\label{sec:mismatch}

\begin{itemize}
\item[\textbf{RQ5.}] What evidence and diagnostic value arise from validating the same property by both formal proof and PBT?
\end{itemize}

\begin{table}[t]
\centering
\caption{Cross-validation of \textsc{Template} proofs and \textsc{Template} PBTs across 400 properties, with each PBT run for 20 test executions.}
\label{tab:cross-val}
\footnotesize
\setlength{\tabcolsep}{4pt}
\setlength{\aboverulesep}{1pt}
\setlength{\belowrulesep}{1pt}
\begin{tabular}{lrrr|r}
\toprule
& \multicolumn{2}{c}{\textbf{Faithful PBT}} & \multirow{2}{*}{\textbf{No faithful PBT}} & \multirow{2}{*}{\textbf{Total}} \\
\cmidrule(lr){2-3}
& \textbf{Passing} & \textbf{Failing} & & \\
\midrule
\textbf{Successful proof}    & 130 & 1  & 5  & 136 \\
\textbf{No successful proof} & 251 &  5 &  8 & 264 \\
\midrule
\textbf{Total}            & 381 &  6 & 13 & 400 \\
\bottomrule
\end{tabular}
\vspace{-0.5em}
\end{table}

We cross-validate the proof and PBT results for all 400 properties under the \textsc{Template} setting, using the artifact classifications from RQ2 and RQ3. Each PBT is executed 20 times. A faithful PBT is failing if at least one execution raises an assertion error and passing otherwise. Table~\ref{tab:cross-val} summarizes the resulting proof and PBT outcomes. 
For \textbf{130}/400 properties, both tracks provide supporting evidence: Lean establishes the intended property over the formal model, while a faithful PBT executes on PySpark without finding a counterexample.
These cases provide the strongest evidence available from the two tracks.

\paragraph{Counterexamples are diagnostic}
A faithful PBT finds a counterexample for \textbf{6} properties, refuting each on the real implementation. For one, a Lean proof also succeeds; the disagreement, invisible to either track alone, exposes a model–runtime gap. Lean proves \texttt{size(array\_except} \texttt{(filter(arr, }$x\to x<0$\texttt{), arr)) = 0} over the total array type \texttt{List}, which has no null inhabitant, while PySpark admits nullable array columns for which the equality fails. This case reveals a gap in the model’s treatment of nullable arrays.

\paragraph{PBT localizes where proofs must grow}
For \textbf{251} properties, a faithful PBT executes without an assertion error, but the proof track produces no successful proof.
Of these, \textbf{107} fall outside the current model. Extending the array model would bring the most into scope---\textbf{58} properties, led by count aggregates (15), \texttt{arrays\_zip} (13), and array reductions (12)---followed by DataFrame extensions such as join, set, and window semantics.
Another \textbf{138} exhaust the proof budget, motivating stronger search such as helper-lemma synthesis~\cite{sivaraman2022data}.
The last \textbf{6} compile but prove nothing of interest.

\vspace{.5em}
\simplebox{black!5}{%
\noindent\textbf{Answer to RQ5.}
For 130 of 400 properties (32.5\%), a Lean proof and a faithful passing PBT agree---the strongest combined evidence. Disagreements are diagnostic: a counterexample to a proven property exposes a model--runtime gap and its fix, and a proof-less passing test localizes where formalization and proof search pay off most.
}
\section{Threats to Validity}

\noindent\textbf{Measurement.}
Assessing candidate-property correctness, proof hallucinations, and PBT faithfulness requires manual semantic judgment and may introduce reviewer error. We release all proofs and tests for independent re-examination. Our coverage metrics measure behavioral breadth rather than semantic adequacy or fault-detection effectiveness.

\noindent\textbf{Experimental design.}
Both synthesis tracks are stochastic and use one run per property and configuration. We compare configurations over the same property sets under fixed model settings, budgets, and environments, and aggregate results across hundreds of properties, reducing sensitivity to isolated generations. Results may still vary across runs and with different template designs, generators, lemmas, or formal models.

\noindent\textbf{Generalizability.}
We evaluate four recurring property families in PySpark using one model configuration. The families represent several forms of equivalence and other relational properties in data-intensive systems, but effect sizes may differ for other families, systems, formalizations, or models.
\section{Related Work}

\paragraph{Property-based testing and property specification}
Goldstein et al.~\cite{goldsteinPropertyBasedTestingPractice2024} empirically study developers' PBT experience and find \emph{property specification} a central obstacle: developers struggle both to identify suitable properties and to turn informal intent into executable ones.
Lahiri~\cite{lahiriIntentFormalizationGrand2026b} frames this as a grand challenge for the age of AI agents: formalizing informal intent into checkable specifications is what makes agent-generated code trustworthy rather than merely abundant.
Hughes et al.~\cite{hughesHowSpecifyIt2019} give a taxonomy of reusable property patterns for pure functions---invariants, postconditions, metamorphic and inductive properties, and model-based specifications.
Segura et al.~\cite{seguraTemplateBasedApproachDescribing2017} represent metamorphic relations as templates that make explicit the source/follow-up inputs and the expected output relation, but use templates primarily as a \emph{documentation} mechanism. Earlier, Dwyer et al.~\cite{dwyerPatternsPropertySpecifications1999} catalog recurring \emph{specification patterns} for temporal properties in finite-state verification, mapping common requirements onto temporal logics for model checking.
Most closely related, agentic tools drive LLMs to infer and run property-based tests: Agentic PBT~\cite{maazAgenticPropertyBasedTesting2025a} finds real bugs across the Python ecosystem, and AWS's Kiro~\cite{kiroCorrectnessPropertybasedTests2025} turns requirements into PBT for ``spec correctness'' in an IDE, while stopping short of formal verification.
Our work is complementary: rather than inferring or generating tests one module at a time, we organize \emph{recurring property families} as templates and validate each instance at scale along \emph{both} a proof track and a PBT track.

\paragraph{Structure-guided synthesis and theorem proving}
A recurring idea in synthesis and theorem proving is to supply partial structure that guides search toward a target \emph{fixed in advance}. \textsc{Sketch}~\cite{solar-lezamaCombinatorialSketchingFinite2006} fills the holes of a partial program against a specification, and DSP~\cite{jiangDraftSketchProve2023} maps an informal proof into a formal sketch that guides an automated prover over easier subproblems. Closest to our setting, SITA~\cite{liSITAFrameworkStructuretoInstance2026} abstracts existing Lean formalizations into reusable structures that an LLM instantiates for concrete theorems---much as our templates parameterize a proof's shared structure. A related line instead supplies the auxiliary facts a proof needs: synthesizing the helper or implication lemmas witnessed during a stuck proof~\cite{yangLemmaSynthesisAutomating2019,brendel2025synthesizing}, or discovering new lemmas by instantiating user-provided schemes (IsaScheme~\cite{montano-rivasSchemebasedTheoremDiscovery2012}) or LLM-generated lemma templates (Lemmanaid~\cite{alhessi2025lemmanaid}), filtering false conjectures by counterexample. Our property templates share this structure-guided view but differ in two ways. First, whereas these approaches synthesize programs or prove mathematical theorems, a single template drives \emph{both} a Lean proof and an executable PySpark test, validating the correctness of \emph{real software systems} by proof and execution alike. Second, these methods attach no precondition to their holes, so an instantiation is a \emph{conjecture} checked after the fact---filtered by counterexample search and proved one at a time---whereas our template carries the family's \emph{local law} (\textsc{AggDecomp}'s decomposition law, \textsc{UDF}'s pointwise equivalence) as a precondition: proved once, \emph{any} satisfying instantiation lifts to a correct property over an arbitrary pipeline and workload, so the agent discharges only the local law and the instance is correct by construction.

\paragraph{Testing and verifying data-intensive and query-processing systems}
A body of work targets the correctness of data-processing systems themselves. On the verification side, automated SQL equivalence provers decide whether two relational queries are equivalent and thereby verify rewrite rules, as in Cosette~\cite{chuCosetteAutomatedProver2017} and SQLSolver~\cite{dingProvingQueryEquivalence2023}; WeTune~\cite{wangWeTuneAutomaticDiscovery2022} goes further, automatically \emph{discovering} new rewrite rules and verifying them with such a prover. On the testing side, SQLancer detects logic and optimization bugs in database engines through constructed oracles such as query partitioning and a non-optimizing reference engine~\cite{riggerFindingBugsDatabase2020b,riggerDetectingOptimizationBugs2020}. Closest to our setting, big-data testers generate Spark inputs by symbolic execution of dataflow operators and UDFs~\cite{gulzarWhiteBoxTestingBig2019} or framework-abstraction fuzzing~\cite{zhangBigFuzzEfficientFuzz2020}. Each fixes both target and technique---one query pair, one engine, or one program at a time. Our work instead treats the optimization-relevant equivalences of data-intensive computing as recurring \emph{property families}, and validates each instance at scale along \emph{both} a proof track and a PBT track over real PySpark.

\section{Conclusion}

We set out to make the validation of a software system’s many correctness properties tractable at scale. Our central idea is to capture each recurring property family once as a \emph{property template}. Each candidate property is then validated along two complementary tracks: a machine-checked Lean~4 proof and an executable property-based test against the real system. In both tracks, the template fixes the shared structure and leaves only property-specific holes to be filled. Instantiated for data-intensive computing on Apache Spark, property templates increase machine-checked synthesis successes by up to $2.6\times$ and reduce proof hallucinations by $59\%$. They also reduce intent misalignments in synthesized tests from 22 to 1, while lowering synthesis cost by up to $5.7\times$.

This experience also surfaces a caution as agentic theorem proving attracts growing attention. A proof accepted by Lean establishes the encoded theorem under its stated definitions and assumptions, but the encoded theorem may not faithfully express the intended property. An agent may misstate the property, introduce vacuous hypotheses, or rely on additional axioms that weaken the intended guarantee. Templates reduce such failures by fixing a family's statement structure, but confirming that a machine-checked proof reflects genuine intent still requires human inspection. Detecting this \emph{formalization gaming}~\cite{kimLLMsGameFormalization2026} automatically\textemdash auditing definitions for faithfulness and proofs for unsound dependencies\textemdash is an important open problem for trustworthy AI-assisted verification.

\section*{Data Availability}
Our code, models, and data are available at~\url{https://anonymous.4open.science/r/AgentLeanDiscprop-1597/}.
\clearpage

\bibliographystyle{IEEEtran}
\bibliography{reference}

@inproceedings{10.1007/978-3-319-22102-1_22,
  title = {Foundational Property-Based Testing},
  booktitle = {Interactive Theorem Proving},
  author = {Paraskevopoulou, Zoe and Hri{\c T}cu, C{\u a}t{\u a}lin and D{\'e}n{\`e}s, Maxime and Lampropoulos, Leonidas and Pierce, Benjamin C.},
  editor = {Urban, Christian and Zhang, Xingyuan},
  year = 2015,
  pages = {325--343},
  publisher = {Springer International Publishing},
  address = {Cham},
  isbn = {978-3-319-22102-1}
}

@inproceedings{chuCosetteAutomatedProver2017,
  title = {Cosette: {{An Automated Prover}} for {{SQL}}},
  booktitle = {8th {{Biennial Conference}} on {{Innovative Data Systems Research}} ({{CIDR}})},
  author = {Chu, Shumo and Wang, Chenglong and Weitz, Konstantin and Cheung, Alvin},
  year = 2017
}

@inproceedings{claessenQuickCheckLightweightTool2000,
  title = {{{QuickCheck}}: A Lightweight Tool for Random Testing of {{Haskell}} Programs},
  booktitle = {Proceedings of the {{Fifth ACM SIGPLAN International Conference}} on {{Functional Programming}} ({{ICFP}} '00)},
  author = {Claessen, Koen and Hughes, John},
  year = 2000,
  pages = {268--279},
  publisher = {ACM},
  doi = {10.1145/351240.351266}
}

@inproceedings{demouraLeanTheoremProver2015,
  title = {The {{Lean Theorem Prover}} ({{System Description}})},
  booktitle = {Automated {{Deduction}} - {{CADE-25}}},
  author = {{de Moura}, Leonardo and Kong, Soonho and Avigad, Jeremy and {van Doorn}, Floris and {von Raumer}, Jakob},
  editor = {Felty, Amy P. and Middeldorp, Aart},
  year = 2015,
  pages = {378--388},
  publisher = {Springer International Publishing},
  address = {Cham},
  doi = {10.1007/978-3-319-21401-6_26},
  isbn = {978-3-319-21401-6},
  langid = {english}
}

@article{dingProvingQueryEquivalence2023,
  title = {Proving {{Query Equivalence Using Linear Integer Arithmetic}}},
  author = {Ding, Haoran and Wang, Zhaoguo and Yang, Yicun and Zhang, Dexin and Xu, Zhenglin and Chen, Haibo and Piskac, Ruzica and Li, Jinyang},
  year = 2023,
  journal = {Proceedings of the ACM on Management of Data},
  volume = {1},
  number = {4},
  pages = {1--26},
  doi = {10.1145/3626768}
}

@inproceedings{dwyerPatternsPropertySpecifications1999,
  title = {Patterns in {{Property Specifications}} for {{Finite-State Verification}}},
  booktitle = {Proceedings of the 21st {{International Conference}} on {{Software Engineering}} ({{ICSE}})},
  author = {Dwyer, Matthew B. and Avrunin, George S. and Corbett, James C.},
  year = 1999,
  pages = {411--420},
  publisher = {ACM},
  address = {New York, NY, USA},
  doi = {10.1145/302405.302672}
}

@inproceedings{goldsteinPropertyBasedTestingPractice2024,
  title = {Property-{{Based Testing}} in {{Practice}}},
  booktitle = {Proceedings of the {{IEEE}}/{{ACM}} 46th {{International Conference}} on {{Software Engineering}}},
  author = {Goldstein, Harrison and Cutler, Joseph W. and Dickstein, Daniel and Pierce, Benjamin C. and Head, Andrew},
  year = 2024,
  month = apr,
  series = {{{ICSE}} '24},
  pages = {1--13},
  publisher = {Association for Computing Machinery},
  address = {New York, NY, USA},
  doi = {10.1145/3597503.3639581},
  urldate = {2026-06-18},
  isbn = {979-8-4007-0217-4}
}

@inproceedings{gulzarWhiteBoxTestingBig2019,
  title = {White-{{Box Testing}} of {{Big Data Analytics}} with {{Complex User-Defined Functions}}},
  booktitle = {Proceedings of the 2019 27th {{ACM Joint Meeting}} on {{European Software Engineering Conference}} and {{Symposium}} on the {{Foundations}} of {{Software Engineering}}},
  author = {Gulzar, Muhammad Ali and Mardani, Shaghayegh and Musuvathi, Madanlal and Kim, Miryung},
  year = 2019,
  pages = {290--301},
  publisher = {ACM},
  address = {New York, NY, USA},
  doi = {10.1145/3338906.3338953}
}

@inproceedings{hughesExperiencesQuickCheck2016,
  title = {Experiences with {{QuickCheck}}: {{Testing}} the {{Hard Stuff}} and {{Staying Sane}}},
  booktitle = {A {{List}} of {{Successes That Can Change}} the {{World}}: {{Essays Dedicated}} to {{Philip Wadler}} on the {{Occasion}} of {{His}} 60th {{Birthday}} ({{LNCS}} 9600)},
  author = {Hughes, John},
  year = 2016,
  pages = {169--186},
  publisher = {Springer},
  doi = {10.1007/978-3-319-30936-1_9}
}

@inproceedings{hughesHowSpecifyIt2019,
  title = {How to {{Specify It}}! {{A Guide}} to {{Writing Properties}} of {{Pure Functions}}},
  booktitle = {Trends in {{Functional Programming}}: 20th {{International Symposium}}, {{TFP}} 2019, {{Vancouver}}, {{BC}}, {{Canada}}, {{June}} 12--14, 2019, {{Revised Selected Papers}}},
  author = {Hughes, John},
  year = 2019,
  month = jun,
  pages = {58--83},
  publisher = {Springer-Verlag},
  address = {Berlin, Heidelberg},
  doi = {10.1007/978-3-030-47147-7_4},
  urldate = {2026-06-13},
  isbn = {978-3-030-47146-0}
}

@inproceedings{jiangDraftSketchProve2023,
  title = {Draft, {{Sketch}}, and {{Prove}}: {{Guiding Formal Theorem Provers}} with {{Informal Proofs}}},
  booktitle = {The {{Eleventh International Conference}} on {{Learning Representations}} ({{ICLR}} 2023)},
  author = {Jiang, Albert Q. and Welleck, Sean and Zhou, Jin Peng and Li, Wenda and Liu, Jiacheng and Jamnik, Mateja and Lacroix, Timoth{\'e}e and Wu, Yuhuai and Lample, Guillaume},
  year = 2023,
  eprint = {2210.12283},
  archiveprefix = {arXiv}
}

@misc{kimLLMsGameFormalization2026,
  title = {Do {{LLMs Game Formalization}}? {{Evaluating Faithfulness}} in {{Logical Reasoning}}},
  author = {Kim, Kyuhee and Poiroux, Auguste and Bosselut, Antoine},
  year = 2026,
  number = {arXiv:2604.19459},
  eprint = {2604.19459},
  publisher = {arXiv},
  doi = {10.48550/arXiv.2604.19459},
  archiveprefix = {arXiv}
}

@misc{kiroCorrectnessPropertybasedTests2025,
  title = {Correctness with {{Property-based}} Tests},
  author = {{Kiro}},
  year = 2025,
  month = nov,
  journal = {Kiro Documentation},
  publisher = {Amazon Web Services},
  urldate = {2026-06-27},
  howpublished = {https://kiro.dev/docs/specs/correctness/},
  langid = {english}
}

@inproceedings{kleinSeL4FormalVerification2009a,
  title = {{{seL4}}: {{Formal Verification}} of an {{OS Kernel}}},
  booktitle = {Proceedings of the {{ACM SIGOPS}} 22nd {{Symposium}} on {{Operating Systems Principles}} ({{SOSP}})},
  author = {Klein, Gerwin and Elphinstone, Kevin and Heiser, Gernot and Andronick, June and Cock, David and Derrin, Philip and Elkaduwe, Dhammika and Engelhardt, Kai and Kolanski, Rafal and Norrish, Michael and Sewell, Thomas and Tuch, Harvey and Winwood, Simon},
  year = 2009,
  pages = {207--220},
  publisher = {ACM},
  doi = {10.1145/1629575.1629596}
}

@misc{lahiriIntentFormalizationGrand2026b,
  title = {Intent {{Formalization}}: {{A Grand Challenge}} for {{Reliable Coding}} in the {{Age}} of {{AI Agents}}},
  author = {Lahiri, Shuvendu K.},
  year = 2026,
  number = {arXiv:2603.17150},
  eprint = {2603.17150},
  publisher = {arXiv},
  doi = {10.48550/arXiv.2603.17150},
  archiveprefix = {arXiv}
}

@article{leroyFormalVerificationRealistic2009,
  title = {Formal Verification of a Realistic Compiler},
  author = {Leroy, Xavier},
  year = 2009,
  journal = {Communications of the ACM},
  volume = {52},
  number = {7},
  pages = {107--115},
  doi = {10.1145/1538788.1538814}
}

@inproceedings{liSITAFrameworkStructuretoInstance2026,
  title = {{{SITA}}: {{A Framework}} for {{Structure-to-Instance Theorem Autoformalization}}},
  booktitle = {Proceedings of the {{AAAI Conference}} on {{Artificial Intelligence}} ({{AAAI}} 2026)},
  author = {Li, Chenyi and Ma, Wanli and Wang, Zichen and Wen, Zaiwen},
  year = 2026,
  volume = {40},
  eprint = {2511.10356},
  pages = {19224--19232},
  publisher = {AAAI Press},
  doi = {10.1609/aaai.v40i23.38997},
  archiveprefix = {arXiv}
}

@misc{maazAgenticPropertyBasedTesting2025a,
  title = {Agentic {{Property-Based Testing}}: {{Finding Bugs Across}} the {{Python Ecosystem}}},
  author = {Maaz, Muhammad and DeVoe, Liam and {Hatfield-Dodds}, Zac and Carlini, Nicholas},
  year = 2025,
  number = {arXiv:2510.09907},
  eprint = {2510.09907},
  publisher = {arXiv},
  doi = {10.48550/arXiv.2510.09907},
  archiveprefix = {arXiv}
}

@article{mcnemarNoteSamplingError1947,
  title = {Note on the Sampling Error of the Difference between Correlated Proportions or Percentages},
  author = {McNemar, Quinn},
  year = 1947,
  month = jun,
  journal = {Psychometrika},
  volume = {12},
  number = {2},
  pages = {153--157},
  issn = {1860-0980},
  doi = {10.1007/BF02295996}
}

@article{montano-rivasSchemebasedTheoremDiscovery2012,
  title = {Scheme-Based Theorem Discovery and Concept Invention},
  author = {{Monta{\~n}o-Rivas}, Omar and McCasland, Roy L. and Dixon, Lucas and Bundy, Alan},
  year = 2012,
  journal = {Expert Systems with Applications},
  volume = {39},
  number = {2},
  pages = {1637--1646},
  doi = {10.1016/j.eswa.2011.06.055}
}

@inproceedings{riggerDetectingOptimizationBugs2020,
  title = {Detecting {{Optimization Bugs}} in {{Database Engines}} via {{Non-Optimizing Reference Engine Construction}}},
  booktitle = {Proceedings of the 28th {{ACM Joint Meeting}} on {{European Software Engineering Conference}} and {{Symposium}} on the {{Foundations}} of {{Software Engineering}}},
  author = {Rigger, Manuel and Su, Zhendong},
  year = 2020,
  pages = {1140--1152},
  publisher = {ACM},
  address = {New York, NY, USA},
  doi = {10.1145/3368089.3409710}
}

@article{riggerFindingBugsDatabase2020b,
  title = {Finding {{Bugs}} in {{Database Systems}} via {{Query Partitioning}}},
  author = {Rigger, Manuel and Su, Zhendong},
  year = 2020,
  journal = {Proceedings of the ACM on Programming Languages},
  volume = {4},
  number = {OOPSLA},
  pages = {1--30},
  doi = {10.1145/3428279}
}

@inproceedings{seguraTemplateBasedApproachDescribing2017,
  title = {A {{Template-Based Approach}} to {{Describing Metamorphic Relations}}},
  booktitle = {2017 {{IEEE}}/{{ACM}} 2nd {{International Workshop}} on {{Metamorphic Testing}} ({{MET}})},
  author = {Segura, Sergio and Dur{\'a}n, Amador and Troya, Javier and Ruiz Cort{\'e}s, Antonio},
  year = 2017,
  pages = {3--9},
  publisher = {IEEE},
  doi = {10.1109/MET.2017.3}
}

@inproceedings{solar-lezamaCombinatorialSketchingFinite2006,
  title = {Combinatorial Sketching for Finite Programs},
  booktitle = {Proceedings of the 12th {{International Conference}} on {{Architectural Support}} for {{Programming Languages}} and {{Operating Systems}} ({{ASPLOS XII}})},
  author = {{Solar-Lezama}, Armando and Tancau, Liviu and Bod{\'i}k, Rastislav and Seshia, Sanjit A. and Saraswat, Vijay A.},
  year = 2006,
  pages = {404--415},
  publisher = {ACM},
  doi = {10.1145/1168857.1168907}
}

@inproceedings{wangWeTuneAutomaticDiscovery2022,
  title = {{{WeTune}}: {{Automatic Discovery}} and {{Verification}} of {{Query Rewrite Rules}}},
  booktitle = {Proceedings of the 2022 {{International Conference}} on {{Management}} of {{Data}}},
  author = {Wang, Zhaoguo and Zhou, Zhou and Yang, Yicun and Ding, Haoran and Hu, Gansen and Ding, Ding and Tang, Chuzhe and Chen, Haibo and Li, Jinyang},
  year = 2022,
  pages = {94--107},
  publisher = {ACM},
  address = {New York, NY, USA},
  doi = {10.1145/3514221.3526125}
}

@inproceedings{yangLemmaSynthesisAutomating2019,
  title = {Lemma {{Synthesis}} for {{Automating Induction}} over {{Algebraic Data Types}}},
  booktitle = {Principles and {{Practice}} of {{Constraint Programming}} ({{CP}} 2019)},
  author = {Yang, Weikun and Fedyukovich, Grigory and Gupta, Aarti},
  year = 2019,
  series = {Lecture {{Notes}} in {{Computer Science}}},
  pages = {600--617},
  publisher = {Springer},
  doi = {10.1007/978-3-030-30048-7_35}
}

@article{zahariaApacheSparkUnified2016,
  title = {Apache {{Spark}}: A Unified Engine for Big Data Processing},
  author = {Zaharia, Matei and Xin, Reynold S. and Wendell, Patrick and Das, Tathagata and Armbrust, Michael and Dave, Ankur and Meng, Xiangrui and Rosen, Josh and Venkataraman, Shivaram and Franklin, Michael J. and Ghodsi, Ali and Gonzalez, Joseph and Shenker, Scott and Stoica, Ion},
  year = 2016,
  journal = {Communications of the ACM},
  volume = {59},
  number = {11},
  pages = {56--65},
  doi = {10.1145/2934664}
}

@inproceedings{zhangBigFuzzEfficientFuzz2020,
  title = {{{BigFuzz}}: {{Efficient Fuzz Testing}} for {{Data Analytics Using Framework Abstraction}}},
  booktitle = {Proceedings of the 35th {{IEEE}}/{{ACM International Conference}} on {{Automated Software Engineering}}},
  author = {Zhang, Qian and Wang, Jiyuan and Gulzar, Muhammad Ali and Padhye, Rohan and Kim, Miryung},
  year = 2020,
  pages = {722--733},
  publisher = {ACM},
  address = {New York, NY, USA},
  doi = {10.1145/3324884.3416641}
}

@article{sivaraman2022data,
  title={Data-driven lemma synthesis for interactive proofs},
  author={Sivaraman, Aishwarya and Sanchez-Stern, Alex and Chen, Bretton and Lerner, Sorin and Millstein, Todd},
  journal={Proceedings of the ACM on Programming Languages},
  volume={6},
  number={OOPSLA2},
  pages={505--531},
  year={2022},
  publisher={ACM New York, NY, USA}
}

@misc{cometfuzz_github,
  author       = {{Apache DataFusion Comet Developers}},
  title        = {{Apache DataFusion Comet: Fuzz Testing}},
  howpublished = {\url{https://github.com/apache/datafusion-comet/tree/03e833b955d369f994d9652026ca3c1eb641acac/fuzz-testing}},
}

@article{brendel2025synthesizing,
  title={Synthesizing Implication Lemmas for Interactive Theorem Proving},
  author={Brendel, Ana and Sivaraman, Aishwarya and Millstein, Todd},
  journal={Proceedings of the ACM on Programming Languages},
  volume={9},
  number={OOPSLA2},
  pages={2254--2278},
  year={2025},
  publisher={ACM New York, NY, USA}
}

@article{alhessi2025lemmanaid,
  title={Lemmanaid: Neuro-Symbolic Lemma Conjecturing},
  author={Alhessi, Yousef and Einarsd{\'o}ttir, S{\'o}lr{\'u}n Halla and Granberry, George and First, Emily and Johansson, Moa and Lerner, Sorin and Smallbone, Nicholas},
  journal={arXiv preprint arXiv:2504.04942},
  year={2025}
}
\end{document}